\documentclass[twocolumn,floatfix,superscriptaddress,aps,10pt,nofootinbib]{revtex4-2}
\usepackage{times}
\usepackage{amssymb}
\usepackage{color}
\usepackage{amsmath}
\usepackage{amsbsy}
\usepackage{amsthm}
\usepackage{graphicx}
\usepackage{bbm}
\usepackage{bm}
\usepackage{epsfig}
\usepackage{xfrac}
\usepackage{xcolor}
\usepackage{enumerate}
\usepackage{multirow}

\definecolor{myurlcolor}{rgb}{0,0,0.7}
\definecolor{myrefcolor}{rgb}{0.8,0,0}
\usepackage[unicode=true,pdfusetitle, bookmarks=true,bookmarksnumbered=false,bookmarksopen=false, breaklinks=false,pdfborder={0 0 0},backref=false,colorlinks=true, linkcolor=myrefcolor,citecolor=myurlcolor,urlcolor=myurlcolor]{hyperref}

\newcommand{\ket}[1]{\left| {#1} \right\rangle}
\newcommand{\bra}[1]{\left\langle {#1}\right|}
\newcommand{\ketbra}[2]{\ket{#1}\!\bra{#2}}
\newcommand{\braket}[2]{\langle #1 | #2 \rangle}

\renewcommand{\t}[1]{\textrm{#1}}
\newcommand{\tr}[0]{\mathrm{Tr}}

\newcommand{\bx}{{\boldsymbol{x}}}

\newcommand{\var}{\theta}
\newcommand{\dvar}{\phi}
\newcommand{\bvar}{{\boldsymbol{\var}}}
\newcommand{\avar}{\varphi}
\newcommand{\bavar}{{\boldsymbol \avar}}
\newcommand{\bmu}{{\boldsymbol \mu}}
\newcommand{\bbeta}{{\boldsymbol \mu}}
\newcommand{\bdvar}{{\boldsymbol \dvar}}
\newcommand{\bayes}{{\rm bayesian}}
\newcommand{\minimax}{{\rm minimax}}

\newcommand{\ins}{\text{in}}

\newcommand{\thmref}[1]{\hyperref[#1]{Theorem~\ref{#1}}}
\newcommand{\lemmaref}[1]{\hyperref[#1]{Lemma~\ref{#1}}}
\newcommand{\figref}[1]{\hyperref[#1]{Fig.~\ref{#1}}}
\newcommand{\figaref}[1]{\hyperref[#1]{Fig.~\ref{#1}a}}
\newcommand{\figbref}[1]{\hyperref[#1]{Fig.~\ref{#1}b}}
\newcommand{\figcref}[1]{\hyperref[#1]{Fig.~\ref{#1}c}}
\renewcommand{\eqref}[1]{\hyperref[#1]{Eq.~(\ref{#1})}}
\newcommand{\eqsref}[2]{\hyperref[#1]{Eqs.~(\ref{#1})-(\ref{#2})}}
\newcommand{\appref}[1]{\hyperref[#1]{Appendix~\ref{#1}} of \cite{SM}}

\newcommand{\mm}{{\Delta^2\tilde\bvar_{\rm minimax}}}

\usepackage{dsfont}

\begin{document}
\title{Multiple-phase quantum interferometry---real and apparent gains of measuring all the phases simultaneously}
\author{Wojciech G{\'{o}}recki}
\affiliation{Faculty of Physics, University of Warsaw, Pasteura 5, 02-093 Warsaw, Poland}
\author{Rafa{\l} Demkowicz-Dobrza{\'n}ski}
\affiliation{Faculty of Physics, University of Warsaw, Pasteura 5, 02-093 Warsaw, Poland}
\begin{abstract}
 We characterize operationally meaningful quantum gains in a paradigmatic model of lossless multiple-phase interferometry
and stress insufficiency of the analysis based solely on the concept
of quantum Fisher information. We show that
the advantage of the optimal simultaneous estimation scheme amounts to a constant factor improvement when
compared with schemes where each phase is estimated separately---contrary to a widely cited results claiming a
better precision scaling in terms of the number of phases involved.
\end{abstract}

\maketitle
%

\paragraph*{Introduction and the summary of results.}
Quantum metrology aims at identifying optimal ways of utilizing quantum systems as sensing probes~\cite{giovannetti2006quantum,Paris2009,giovannetti2011advances,Toth2014,Demkowicz2015,Schnabel2016,degen2017quantum,Pezze2018,Pirandola2018}. When $N$ quantum probes are used independently, estimation variance decreases inversely proportional to the number of probes---the standard quantum limit (SQL). The hallmark of quantum metrology is the potential quadratic scaling improvement over the SQL---the Heisenberg limit (HL) \cite{caves1981quantum,holland1993interferometric,lee2002quantum,wineland1992spin,mckenzie2002experimental,bollinger1996optimal,leibfried2004toward,giovannetti2004quantum,huelga1997improvement,de2005quantum,pezze2009}.

In \emph{multiple}-parameter estimation scenarios, simultaneous estimation of $p$ parameters
in a single experiment may additionally provide an improved performance when compared with strategies where each parameter is estimated separately \cite{ragy2016, baumgratz2016, Yuan2016, tsang2016quantum, Liu2017, Nichols2018, Gorecki2020}.



\begin{figure}[hbt!]
\includegraphics[width=0.8\columnwidth]{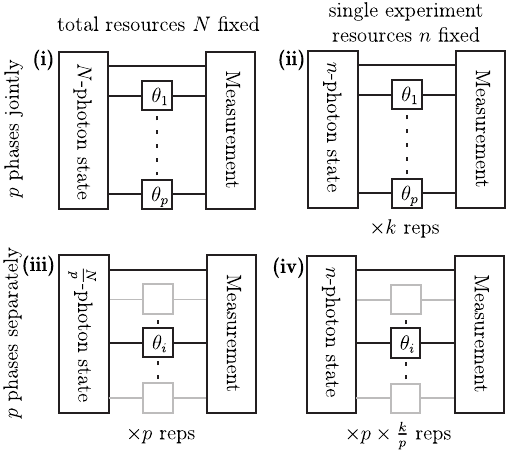}
\caption{Multiple-phase estimation schemes, where a constraint is imposed on the total resources used (left column) or
resources used in a single experiment (right column). The top and bottom rows represent respectively the protocols where all the phases are measured jointly or the estimation procedure is repeated for each of the phases separately.}
\label{fig:scheme}
\end{figure}

A paradigmatic model to study the potential of multi-parameter quantum enhanced metrological protocols is
the multiple-phase estimation problem. The goal is to estimate all the relative phase shifts in a multiple-arm interferometer with the best precision possible given a constraint on the total number of photons used,
see Fig.~\ref{fig:scheme}.

The most common tool to analyze the potential of quantum metrological strategies is the Quantum Fisher Information (QFI),
inverse of which lower-bounds the variance of any locally unbiased estimator $\tilde{\theta}$ via the famous quantum Cramer-Rao (CR)
bound \cite{helstrom1976quantum, Holevo1982, braunstein1994statistical}. In the single parameter case the CR bound takes the form:
\begin{equation}
\Delta^2 \tilde{\var} \geq \frac{1}{k F\left(\rho^{n}_\var\right)},
\end{equation}
\begin{table}[hbt!]
\begin{tabular}{|c|c|c|c|}
\hline
\multirow{3}{6em}{$\quad \Delta^2\tilde{\bvar} \geq$} & \multirow{3}{5em}{\centering{Independent}  \\ probes \\ \vspace{0.2em} SQL }  & \multicolumn{2}{|c|}{Entangled probes}  \\
\cline{3-4}
    &   & \multirow{2}{7em}{$N= k n $ probes \\ \centering{HL}} & \multirow{2}{7em}{\centering{$n$ probes, $k$ reps}\\  HS}
\\ & & &
\\
  \hline
\centering{single phase}  & \phantom{${\displaystyle \int}$}\!\!\!\! $\dfrac{1}{N}$&  $\dfrac{\pi^2}{N^2}$  & $\dfrac{1}{k n^2}$\\
\hline
\multirow{2}{6em}{\centering{$p$-phases}\\ \centering{jointly}} & \multirow{2}{5em}{\centering{$\dfrac{p^2}{4 N} $}}  &
\multirow{2}{6em}{\!\!\!\! {\Large $^{\textrm{(i)}}$} \quad \    $\dfrac{\boldsymbol{c \,p^3}}{\boldsymbol{N^2}}$} & \multirow{2}{5em}{\!\!\!\!\!\!\! {\Large $^{\textrm{(ii)}}$} \ \  $\dfrac{{p^2}}{{4 k n^2}}$}\\
 & & &\\
 \hline
 \multirow{2}{6em}{\centering{$p$-phases}\\ \centering{separately}} & \multirow{2}{5em}{\centering{$\dfrac{p^2}{N}$}} &  \multirow{2}{5em}{\!\!\!\!\!\!\! {\Large $^{\textrm{(iii)}}$} \ \  {$\dfrac{\pi^2 p^3}{N^2}$}}  & \multirow{2}{5em}{\!\!\!\!\!\!\! {\Large $^{\textrm{(iv)}}$} \ \  {$\dfrac{p^2}{k n^2}$}}\\
 & & &\\
\hline
\end{tabular}
\caption{Asymptotically achievable lower bounds on the sum of variances of estimated phases. 
The main result of the paper is the bolded formula representing the proper mutiple-phase Heisenberg limit (HL) and demonstrating the $p^3$ scaling with the number of estimated phases.
All bounds are tight in the asymptotic limit $N \rightarrow \infty $ (or $k \rightarrow \infty$) while in case of joint phases estimation tightness requires an additional $p \rightarrow \infty$ limit
 ($c=1.89$ yields a universally valid bound, while $c=2$ yields an asymptotically achievable cost).
 The SQL column can be regarded as a special case of the HS column when $n=1$, and $k = N$.}
 \label{tab:results}
\end{table}
where $k$ is the number of repetitions of an experiment, while $F(\rho^n_\var)$
is the QFI computed on the $n$-probe output state on which the parameter $\var$ has been encoded.
In case of the standard two-arm optical interferometry, a single relative phase between the two arms is being estimated,
and $F(\rho^{n}_\var) = n$ for $n$ uncorrelated photons sent into the interferometer, while the maximum value
$F(\rho^{n}_\var) = n^2$ is obtained for optimally entangled state of $n$ photons---the n00n state---resulting
in the $1/n^2$ Heisenberg scaling (HS) of precision \cite{lee2002quantum,wineland1992spin,mckenzie2002experimental,bollinger1996optimal,leibfried2004toward,giovannetti2004quantum,
huelga1997improvement,de2005quantum, Demkowicz2015}.
In general, this bound is operationally saturable provided one takes the asymptotic  limit $k \rightarrow \infty$, while keeping the $n$ fixed.
Such a case corresponds to an experimental realization, where the amount of resources used in a single realization is large but limited,
while experiment may be repeated arbitrary number of times.

However, a fundamental question is  what is the true HL for precision
if the total amount of resources $N = n\cdot k$ is restricted and the $N\to \infty$ limit is taken. Since the scaling of precision is quadratic in $n$ and linear in $k$, the optimal choice appears to be $n=N, k=1$. However, in this case one can not use saturability arguments based on the many-repetition scenario and the predictions of the QFI
may be misleading with respect to the choice of the optimal probe states as well as the asymptotically achievable
precision limit.
This becomes clear when the results are contrasted with the ones obtained
the minimax \cite{hayashi2011}, Bayesian \cite{Gorecki2020pi} or information theoretic approach \cite{Hall2012}.

The use of a single N00N state (note $N$ instead of $n$ is intentional) is clearly not an operationally meaningful strategy
when discussing the HL, as it is not capable of discriminating phases that differ by a multiple of $2\pi/N$.
There is clearly a need to sacrifice part of the resources to get rid of the arising ambiguity, and
this leads to a $\pi^2$ increase in the asymptotically saturable bound, which can be rigorously shown within a Bayesian estimation framework \cite{luis1996,buzek1999,Berry2000,higgins2009demonstrating, Berry2009,Kaftal2014,Gorecki2020pi}.
Therefore, in order to avoid confusion, we will introduce a clear distinction between
the two approaches and refer to them as
\begin{enumerate}
\item[(HS)]{the amount resources used in single repetition of experiment $n$ large  but finite,
and whole experiment may be repeated $k \rightarrow \infty$ times: $\Delta^2\tilde\var \propto \frac{1}{k n^2}$}
\item[(HL)]{the total amount of resources $N$ is restricted, and no repetitions of an experiment are
assumed: $\Delta^2\tilde\var \propto \frac{1}{N^2}$}.
\end{enumerate}
The two approaches may only be reconciled provided one is able to properly account for the scaling of the
required number of repetitions $k$ with the increasing number of probes $n$ used in a single experiment, that guarantees
saturability within the HS approach. This is, however, hardly ever possible and typically the issue is simply ignored in the literature.


In the multiparameter case a rigorous study of the achievable HL is much more challenging and
the common approach is to rather work in the HS paradigm where efficiently computable multiparameter generalizations of CR bounds are used \cite{matsumoto2002new, Yuan2016,baumgratz2016quantum,Gessner2018,Kura2018,ge2018distributed,Gorecki2020,demkowicz2020multi}.
Rigorous analysis of the actual saturable HL are typically restricted to Bayesian framework case studies utilizing some underlying group structure of the problem \cite{bagan2000,bagan2001, chiribella2004,bagan2004su2,chiribella2005su2,hayashi2006su2, hayashi2016fourier, kahn2007}.
However, quite surprisingly, the actual analytical form of the asymptotically saturable HL for the paradigmatic multiple phase estimation problem is missing (see \cite{pezze2021} for a recent numerical attempt to tackle the problem).

In this paper, we employ an operationally meaningful minimax approach to
derive an asymptotically saturable HL for the multiple-phase estimation problem and demonstrate that it manifests
a $p^3$ scaling, with the number of parameters involved, rather than the $p^2$ that is advocated when following the HS approach \cite{Humphreys2013}. We also clarify apparent gains that may be obtained thanks to simultaneous phase estimation, when compared with strategies that estimate all the phases separately. We show that the advantage amounts to a constant factor gain and,
contrary to the claims of \cite{Humphreys2013,Szczykulska2016, liu2019quantum}, does not lead to a better scaling of precision with the number of parameters involved. We explain the apparent contradiction by pointing out the improper use of saturability arguments
that are often invoked when following the HS approach---Tab.~\ref{tab:results} summarizes the main results of this paper.

\paragraph*{Multiple-phase estimation problem.}
Consider a multiple-phase estimation problem as depicted in Fig.~\ref{fig:scheme}, where
the goal is to estimate the value of $p$-phases $\bvar=[\var_1,\dots,\var_p]$---
relative phase delays in the $i$-th arm of an interferometer with respect to the reference arm.
For a general $n$-photon state at the input, the output state with phase information encoded will have the form:
\begin{equation}
\label{photonstate}
\ket{\Psi_\bvar^n}=\sum_{\boldsymbol{m}:\sum_{i=1}^p m_i \leq n} c_{\boldsymbol{m}}e^{i\boldsymbol{m}\cdot \bvar}\ket{m_1,m_2,...m_p},
\end{equation}
with $\boldsymbol{m} = [m_1,\dots,m_p]$, where $m_i$ represents the number of photons in the $i$-th `signal' arm, while remaining $n - \sum_{i=1}^p m_i$ photons are sent through reference arm.
A general quantum measurement is then performed in order to extract information on the encoded phases,
mathematically specified by a set of positive operators $\{M_{x}\}$, $\sum_x M_x = \openone$, where $x$ labels a measurement outcome
observed with probability $p_{\bvar}(x) = \bra{\Psi_\bvar^n} M_x \ket{\Psi_\bvar^n}$.
The measurement outcome(s) is then fed into an estimator function which yields the inferred values of the phases.
In the scenario (i), the estimator $\tilde{\bvar}(x)$
is a function of just a single measurement outcome (in this case $n=N$ as all resources are used in a single shot);
in (ii) the experiment is repeated $k$ times and the estimator is a function of all $k$ measurement outcomes
$\tilde{\bvar}(x^1,\dots,x^k)$; in (iii) $p$ separate protocols, involving $p$ different states (each containig $N/p$ photons) and different measurements, are performed yielding measurement outcomes $x_1,\dots,x_p$ where each outcome $x_i$ feeds the estimator of the $i$-th phase $\tilde{\var}_i(x_i)$; finally, in (iv) each of the $p$ separate protocol is repeated $\sfrac{k}{p}$ times, yielding in total $p \times \sfrac{k}{p}=k$ measurement outcomes and resulting in $p$ separate estimators of each phase $\tilde{\var}_i(x_i^1,\dots,x_i^{\sfrac{k}{p}})$, $i\in\{1,\dots,p\}$.

Irrespectively of which scenario is considered, the figure of merit to be minimized is the sum of squared errors of estimated phases
\begin{equation}
\label{eq:cost}
\Delta^2\tilde\bvar =\int \t{d}\boldsymbol{x}\, p_{\bvar}(\bx)(\tilde\bvar(\bx)-\bvar)^2,
\end{equation}
where $\int \t{d}\boldsymbol{x}$ formally represents integration over all (possibly continuous) measurement outcomes and $(\tilde\bvar(\bx)-\bvar)^2= \sum_{i=1}^p (\tilde\var_i(\bx)-\var_i)^2$.
As this is a point-wise figure of merit (calculated at a given $\bvar$), in order to make the minimization task meaningful one needs
to impose additional constraints on the estimator function---otherwise a trivial solution $\tilde{\bvar}(\boldsymbol{x}) = \bvar$ yields zero cost.

The most common one is the locally unbiasedness condition, which is also the key assumption
behind the derivation of the CR-type bounds \cite{helstrom1976quantum, Holevo1982, braunstein1994statistical,liu2019quantum, demkowicz2020multi}.
This assumption itself may not be sufficient to obtain an operationally saturable bounds, as in principle the region where the
use of the local-unbiased estimator makes sense may shrink when taking the asymptotic limit $N \rightarrow \infty$ \cite{hayashi2011,Hall2012, Gorecki2020pi}.

Alternatively, one may follow the so called \emph{minimax} approach,
and define a 
region $\Theta$ inside which the true value of $\bvar$ is guaranteed to be and then consider the estimator which gives the best results in the most pessimistic scenario, i.e. which minimizes the cost maximized over all $\bvar\in\Theta$:
\begin{equation}
\label{eq:minmax}
\mm\equiv\inf_{M_{\boldsymbol{x}},\tilde\bvar(\boldsymbol{x})}\sup_{\bvar\in\Theta}\int \t{d}\boldsymbol{x}\, p_{\bvar}(\boldsymbol{x})(\tilde\bvar(\boldsymbol{x})-\bvar)^2.
\end{equation}
The advantage of the approach is that $\Theta$ is fixed while taking the asymptotic limit $N \rightarrow \infty$
and hence no region shrinking issues arise. We now proceed to derive an asymptotically saturable lower bound on the above cost
in the most fundamental scenario (i) and then contrast it with 
scenarios (ii), (iii), (iv).

\paragraph*{Derivation of the multiple-phase HL.} Below we present the sketch of the proof. For more formal derivation see Supplemental Material
 (SM)~\ref{app:A}~\cite{Dominic,Gorecki2020pi,HOLEVO1979385,OZAWA198011,Ozawa1980,bogomolov1982minimax,hayashi2016fourier,Holevo1982,Tsang2020,tanaka2014quantum,boas1995}.
First, we consider an extension of the model by replacing discrete variables $m_i\in\{0,1,...,N\}$ with continuous ones $\frac{m_i}{N}\to\mu_i\in[0,1]$, and the sums by the respective integrals.
Note that such an extension may only decrease the minimal achievable cost, as the discrete model may be always arbitrary well approximated as a special case of the continuous model. The probe-state is now characterized by a $p$-dimensional wave function $f(\boldsymbol{\mu})$:
\begin{equation}
\label{eq:contstate}
\ket{\Psi_{f,\bvar}^N}=\int\displaylimits_{\forall \mu_i\geq 0, \sum_i{\mu_i}\leq 1}d\boldsymbol{\mu}\, f(\boldsymbol{\mu})e^{i N\boldsymbol{\mu} \, \bvar}\ket{\mu_1,\mu_2,..\mu_p}.
\end{equation}

Next, as we argue in detail in the SM, the asymptotic bound for any finite region $\Theta$ is equivalent,
up to the leading $1/N^2$ order, to the cost when the region is unbounded $\Theta = \mathbb{R}^p$. In the latter case
the problem is covariant with respect to the translation group and the optimal measurement
can be restricted to the class of covariant measurements \cite{Holevo1982} (thanks to the generalization of the Hunt-Stein lemma~\cite{HOLEVO1979385,OZAWA198011} for noncompact groups ~\cite{Ozawa1980,bogomolov1982minimax,hayashi2016fourier})
and without loss of generality may be chosen to be
the momentum projection measurement, $M_{\tilde{\bvar}} = \ketbra{\chi_{\tilde\bvar}}{\chi_{\tilde\bvar}}$, where
\begin{equation}
\ket{\chi_{\tilde\bvar}}=\frac{1}{\sqrt{(2\pi/N)^p}}\int \t{d}\boldsymbol{\mu}\, e^{iN\boldsymbol{\mu}\tilde\bvar}\ket{\boldsymbol{\mu}}.
\end{equation}
Note that we have implicitly replaced the measurement outcomes $x$ with the actual estimated values $\tilde{\bvar}$.
Minimization of the resulting lower bound over the probe state wave function $f(\bvar)$ leads to the following 
lower bound on the cost $\Delta^2\tilde\bvar$:

\begin{equation}
\min_f \int\displaylimits_{\mathbb R^p} \t{d} \tilde{\bvar}\, |\braket{\chi_{\tilde{\bvar}}}{{\Psi_{f,\bvar}^N}}|^2 (\tilde\bvar-\bvar)^2  = \frac{1}{N^2}\min_{f}\int\displaylimits_{\mathbb R^p} \t{d} \tilde{\bvar}\,
\left|\hat{f}(\tilde\bvar)\right|^2\tilde\bvar^2
\end{equation}
where $\hat{f}$ is the Fourier transform of $f$ and we dropped the irrelevant dependence on $\bvar$. 
Going back to the $\bmu$-representation, the minimization problem takes the following form:
\begin{equation}
\label{eq:jointenergy}
\begin{split}
&\frac{1}{N^2}\min_{f}\int\displaylimits_{\forall \mu_i\geq 0, \sum_i{\mu_i}\leq 1} \t{d}\boldsymbol{\mu}\, f^*(\boldsymbol{\mu})\left(\sum_{k=1}^p-\partial_{\mu_k}^2\right)f(\boldsymbol{\mu}),\\
&{\rm with}\quad \int\displaylimits_{\forall \mu_i\geq 0, \sum_i{\mu_i}\leq 1} \t{d}\boldsymbol{\mu}\, |f(\boldsymbol{\mu})|^2=1,\\
&f(\boldsymbol{\mu})=0\quad \textrm{for } \boldsymbol{\mu} \textrm{ on the boundary } (\mu_i=0 \lor\sum_i \mu_i=1 ).
\end{split}
\end{equation}
This problem is therefore equivalent to identifying the ground state energy of a quantum particle in a $p$-dimensional simplex-shaped infinite potential well, which in general has not known analytical solution (apart from specific cases \cite{li1984particle,Krishnamurthy_1982,Turner_1984}). 
Still, it may be easily bounded from below in following way.

Since the problem enjoys an inherent symmetry with respect to permuting the $p$ `phase' arms (the reference arm is distinguished by the choice of the cost function), and the total number of photons in $p$ `phase' arms is $\leq N$,
the expectation value of the number of photons in each single `phase arm' is $\leq N/p$.
Now we will neglect the fact, than the distribution of photons in each single arm comes from the multi-arm distribution of $N$ photons and keep only the constraint on the photon expectation value. Such a constraint is a weaker one than the original one, but as it refers just to a single `phase' mode it allows for an effective separation of variables. This allows us to lower bound the total cost by $p$ times
the minimal single-phase estimation cost given mean number of photons in the mode $N/p$:
\begin{equation}
\label{eq:singlemodebound}
\Delta^2\tilde\bvar \geq p \times\frac{1}{N^2}
\min_g \int_0^\infty \t{d}\mu\, g^*(\mu)\left(-\frac{\partial^2}{\partial\mu^2}\right)g(\mu)
\end{equation}
with constraints:
\begin{equation}
\label{cons_main}
 g(0)=0,\ \int_0^\infty \t{d}\mu\, |g(\mu)|^2=1,\  \int_0^\infty \t{d}\mu\, |g(\mu)|^2N\mu =\sfrac{N}{p}.
\end{equation}
The single mode problem may be solved using the standard Lagrange multiplier method, and we obtain the solution $g(\mu)$ to be the Airy function $\t{Ai}(\cdot)$ (see also \cite{Summy1990, Baker2021}, where the same solution appeared in a single phase estimation context),
leading to the final bound
\begin{equation}
\label{eq:minimaxboundfinal}
\Delta^2\tilde\bvar\geq \frac{p^3}{N^2}\frac{4 |A_0|^3}{27}\approx \frac{1.89 p^3}{N^2},
\end{equation}
where $A_0\approx -2.34$ is the first zero of $\t{Ai}(\cdot)$. The most important feature of the bound is the $p^3$ scaling. This bound is valid, even if one considers the most general adaptive strategy 
with arbitrary large ancilla is allowed, see SM~\ref{app:A}.

Note, that an analogue reasoning could not be performed to bound the QFI, as the QFI may be arbitrary large
when only the constraint on the \emph{mean} (and not the maximal) number of photons in the sensing arm is imposed, and leads to some
operationally unjustified claims of sub-Heisenberg estimation strategies \cite{Anisimov2010, Rivas2012}, as discussed in \cite{pezze2013}.

\paragraph*{Comparison of different approaches.}
When following the (ii) approach and minimizing the trace of the inverse of the QFI matrix of the output state, one obtains the following bound on the total cost arising from the applicaton of the multi-parameter version of the CR bound \cite{Humphreys2013} (see also \cite{Yuan2016}
for justification of fundamental optimality):
\begin{equation}
\label{eq:costii}
\Delta^2 \tilde{\bvar} \geq \frac{1}{k}\min_{\ket{\Psi^n}}\tr[F^{-1}(\ket{\Psi^n_\bvar})]=\frac{(1+\sqrt{p})^2p}{ 4 k n^2}\overset{p \gg 1}{\approx} \frac{p^2}{ 4 k n^2},
\end{equation}
where the optimal input state has the form
\begin{equation}
\label{eq:stateii}
\ket{\Psi^n}=\beta\ket{n,0,...,0}+\alpha\left(\ket{0,n,...,0}+...\ket{0,0,...,n}\right),
\end{equation}
with $\alpha=1/\sqrt{p+\sqrt{p}},\beta=1/\sqrt{1+\sqrt{p}}$.
The most visible difference between the two approach is the scaling of the cost with the number of parameters estimated, $p^3$ in (i) vs $p^2$ in (ii).
In order to avoid contradiction, this implies that when considering $k$ repetitions in the (ii) scenario, the actual number of repetitions required to saturate the CR bound will in fact need to increase at least linearly with $p$.
This fact lies at the
heart of the discrepancy between the claims of  \cite{Humphreys2013} and ours.
 Interestingly, when considering the Gaussian states only,
the QFI based study \cite{Gagatsos2016} yields results qualitatively equivalent to ours
($p^3$ cost scaling for both joint and separate strategies) which should be attributed to the fact that the saturability of the CR-type bounds in
Gaussian models is guaranteed already at the single shot level without invoking the multiple repetition argument \cite{Holevo1982, demkowicz2020multi}.

In scenario (iii), one separately sends $N/p$ photons states, into the $i$-th and the reference arm,
in order to measure a given $\theta_i$ phase, using the optimal state for sensing a single completely unknown phase \cite{luis1996,buzek1999,Berry2000}:
\begin{equation}
\label{singleopt}
\ket{\Psi^{\sfrac{N}{p}}_i}= \sqrt{\frac{2}{\sfrac{N}{p}+2}} \sum_{m=0}^{\sfrac{N}{p}}\sin\left[\frac{(m+1)\pi}{\sfrac{N}{p}+2}\right]\ket{\sfrac{N}{p} -m}_0\ket{m}_i,
\end{equation}
where $\ket{\sfrac{N}{p}-m}_0 \ket{m}_i$ denotes a state where $m$ photons is sent into the $i$-th arm and $\sfrac{N}{p} - m$ into the reference arm.
The resulting bound on the total variance is therefore lower bounded by $p$ times the single phase estimation variance \cite{luis1996,buzek1999,Berry2000}:
\begin{equation}
\Delta^2 \tilde\bvar  \overset{N/p \gg 1}{\gtrsim}
 p \times \frac{\pi^2}{(\sfrac{N}{p})^2}=\frac{p^3 \pi^2}{N^2}.
\end{equation}
We see the same scaling as in the joint phase estimation protocol (i), which implies that the largest possible gain coming from joint phase estimation amounts to a constant factor $\leq \pi^2/1.89$. In order to show that the gain over the separate strategy
is indeed achievable, we need to find a state wavefunction $f(\boldsymbol{\mu})$ that manifests an advantage over the separate strategy,
when plugged into the joint estimation cost formula \eqref{eq:jointenergy}. We propose a simple ansatz for the structure of the state that satisfies the boundary conditions
\begin{equation}
\label{eq:state}
f(\boldsymbol{\mu})\propto\left(\prod_{i=1}^p \mu_i\right)^\alpha \left(1-\sum_{i=1}^p \mu_i\right)^\beta.
\end{equation}
The minimal cost
\begin{equation}
\Delta^2\tilde\bvar= \frac{p(1+2\sqrt{p})^2\sqrt{p}(4p+2\sqrt{p}-1)}{(8\sqrt{p}-4)N^2}\overset{p\gg 1}{\approx}\frac{2p^3}{N^2}.
\end{equation}
is obtained for  $\alpha = 3/2,\beta = \sqrt{p}$.
For large $p$ the cost approaches closely the fundamental bound ($2$ vs. $1.89$ coefficient), demonstrating that
the $\pi^2/2 \approx 4.93$ advantage of joint phase estimation over separate strategies is achievable. Note that although this result  was obtained for the problem with continuous variables $\mu_i$, it may be arbitrary well approximated within the original discrete model \eqref{photonstate} with increasing $N$ (in the same spirit as discussed in~\cite{Imai2009} for the single parameter case). See also SM~\ref{app:B}~\cite{luis1996,buzek1999,Berry2000,Imai2009},
for the details of computation, more discussion on the structure of the state and numerical investigation of the convergence of the discrete model to the continuous one when $N$ is being increased.

Finally, the optimal strategy in (iv) is to use subsequently $p$ n00n states
\begin{equation}
\ket{\Psi^n_i} = \frac{1}{\sqrt{2}}\left( \ket{0}_0 \ket{n}_i + \ket{n}_0 \ket{0}_i  \right)
\end{equation}
each designed to sense the $i$-th phase optimally. Since the total number of repetitions is $k$, each phase will be sensed $k/p$ times and hence the final cost resulting from the application of the CR bound reads:
\begin{equation}
\Delta^2\tilde\bvar \geq p \times \frac{1}{\sfrac{k}{p}}\times \frac{1}{n^2} = \frac{p^2}{k n^2}.
\end{equation}
Comparing this result with \eqref{eq:costii} we see that joint phase estimation offers again just a constant factor improvement over separate strategies. This result contrasts the claims of \cite{Humphreys2013}
where a scaling improvement ($p^2$ vs $p^3$) was claimed. Indeed, if  $n/p$ photons were used in a single phase estimation experiment, instead of
considering $k/p$ uses of the $n$-photon state, one would obtain the bound on the cost in the form: $\Delta^2\tilde\var \geq p \times \frac{1}{k} \frac{1}{(\sfrac{n}{p})^2} = \frac{p^3}{k n^2}$.
This latter calculation, however, does not reflects the cost of the optimal separate strategy in the framework that there is some fixed number of photons $n$ used in a single experiment and the experiment is repeated $k$ times.
 The optimal strategy is captured by the former reasoning, leading to the $p^2$ scaling, as we can always regard this strategy as an equivalent $k$ repetitions of an experiment using
a mixed state $\rho^n = \frac{1}{p}\sum_{i=1}^p \ket{\Psi^n_i}\bra{\Psi^n_i}$, where
the factor $\frac{1}{p}$ in formula for $\rho^{n}$ express the fact, that in each repetition we measure only one parameter, with equal probability for each of them. Effectively $k/p$ repetition for each parameter is performed, with $n$ resources each time.

\paragraph*{Conclusions and discussion.}
In this paper we have clarified the relation between formulas for the optimal cost in multiple-phase interferometry obtained within different paradigms, involving either fixing the total resources used or the resources used in a single experiment. Doing so, we have shown that within both paradigms joint phases estimation leads to at most constant factor improvement over the optimal separate strategies.
This constant factor improvement may be attributed to the fact, that in the limit of many phases being sensed, the number of photons needed to be sent into the reference arm becomes negligible
compared with the total number of photons used, whereas in the separate strategy it effectively consumes half of the  resources available (see SM~\ref{app:B} for broader discussion).
This claim remains valid also in the lossy optical interferometry case (where, however, HS does not occurs) as shown in \cite{Albarelli2021}.


Note, that similar issues regarding the apparent scaling advantage of joint vs. separate parameter estimation may arise in other multiparameter estimation problems \cite{Yuan2016, baumgratz2016}
and in order to arrive at operationally meaningful conclusions, one should avoid implicit switching between (i-iv) paradigms and
be aware of non-trivial saturability issues when following the QFI based approach.

\begin{acknowledgments}
We thank Dominic Berry for sharing his argument that we use in Supplemental Material~A1, Francesco Albarelli for numerous fruitful discussions, as well as Howard M. Wiseman, Animesh Datta and Luca Pezz{\'e} for useful comments.
This work was supported by the National Science Center (Poland) grant No.\ 2016/22/E/ST2/00559.
\end{acknowledgments}

\onecolumngrid
\appendix


\renewcommand\thefigure{\thesection\arabic{figure}}    
\setcounter{figure}{0}   

\section{Formal derivation of the bound \eqref{eq:minimaxboundfinal} and its application for local estimation problem}
\label{app:A}
In this appendix we discuss the problem of $p$-phase shift estimation for the continuous model (i.e. after replacing $\frac{m_i}{N}\to \mu_i\in[0,1]$). Moreover, we consider the most general adaptive strategy, when the amount of resources $N$ is the number of application of an elementary gate $U_\bvar$, where arbitrary large ancilla is allowed as well as the action on the state with additional unitaries $V_i$ during the evolution. 

Formally, let us consider $\mathcal H_S=L^2(\{[\mu_1,...,\mu_p]:\forall_i \mu_i\geq 0, \sum_{i=1}^p \mu_1\leq 1\})$ (square-integrable functions on the set $\{[\mu_1,...,\mu_p]:\forall_i \mu_i\geq 0, \sum_{i=1}^p \mu_1\leq 1\}\subset \mathbb R^p$). We define $U_\bvar$ to be a unitary gate, which action on the state $\int\displaylimits_{\forall \mu_i\geq 0, \sum_i{\mu_i}\leq 1} \t{d}\bmu f(\bbeta)\ket{\bbeta}\in\mathcal H_S$ is defined as:
\begin{equation}
U_\bvar \left(\int\displaylimits_{\forall \mu_i\geq 0, \sum_i{\mu_i}\leq 1} \t{d}\bbeta f(\bbeta)\ket{\bbeta}\right)=\int\displaylimits_{\forall \mu_i\geq 0, \sum_i{\mu_i}\leq 1} \t{d}\bbeta e^{i\bbeta\bvar} f(\bbeta)\ket{\bbeta}.
\end{equation}
Consider the most general adaptive strategy, where one chooses an arbitrary initial state $\ket{\Psi_{\rm in}}\in\mathcal H_S\otimes \mathcal H_A$ (where $\mathcal H_A$ may be arbitrary large) and acts on it with a  unitary gate $U_\bvar\otimes \openone_A$ $N$ times and unitary controls $V_i$ in between:
\begin{equation}
\label{generalstate}
\ket{\Psi^N_\bvar}=V_N(U_\bvar\otimes \openone_A) V_{N-1}...V_i(U_\bvar\otimes \openone_A)\ket{\Psi_{\ins}}.
\end{equation}
In particular, if one chooses $\mathcal H_A=\mathbb C$, $V_i=\openone$ and $\ket{\Psi_{\ins}}=\int\displaylimits_{\forall \mu_i\geq 0, \sum_i{\mu_i}\leq 1}\t{d}\bbeta f(\bbeta)\ket{\bbeta}$ they receive the state from \eqref{eq:contstate}:
\begin{equation}
\label{simplystate}
\ket{\Psi^N_\bvar}=\int\displaylimits_{\forall \mu_i\geq 0, \sum_i{\mu_i}\leq 1}\t{d}\bbeta e^{iN\bbeta\bvar}f(\bbeta)\ket{\bbeta}.
\end{equation}

Let us name by $\minimax(\Theta,N)$ the minimax cost for $\bvar\in\Theta$, optimized over initial state, unitary controls during evolution and the measurement:
\begin{equation}
\minimax(\Theta,N)\equiv\inf_{M_{\tilde\bvar},\{V_i\}, \ket{\Psi_{\rm in}}\in\mathcal H_S\otimes \mathcal H_A}\sup_{\bvar\in\Theta}\int \t{d}\tilde\bvar\, \bra{\Psi^N_\bvar}M_{\tilde\bvar}\ket{\Psi^N_\bvar}(\tilde\bvar-\bvar)^2.
\end{equation}
Then the following theorems hold:

\textbf{Theorem 1.} For completely unknown parameters $\Theta=\mathbb R^p$, the optimal cost may be obtained
by applying covariant measurement, without invoking  unitary controls during the evolution  (e.i. $\forall_i V_i=\openone$), resulting in eqref{eq:jointenergy}:
\begin{equation}
\label{thm1}
\begin{split}
\minimax(\mathbb R^p,N)=&\frac{1}{N^2}\min_{f}\int\displaylimits_{\forall \mu_i\geq 0, \sum_i{\mu_i}\leq 1} \t{d}\boldsymbol{\mu}\, f^*(\boldsymbol{\mu})\left(\sum_{k=1}^p-\partial_{\mu_k}^2\right)f(\boldsymbol{\mu}),\\
&{\rm with}\quad \int\displaylimits_{\forall \mu_i\geq 0, \sum_i{\mu_i}\leq 1} \t{d}\boldsymbol{\mu}\, |f(\boldsymbol{\mu})|^2=1,\\
&f(\boldsymbol{\mu})=0\quad \textrm{for } \boldsymbol{\mu} \textrm{ on the boundary } (\mu_i=0 \lor\sum_i \mu_i=1).
\end{split}
\end{equation}

\textbf{Theorem 2.} The $\minimax(\mathbb R^p,N)$ may be bounded from below by \eqref{eq:minimaxboundfinal}:
\begin{equation}
\label{thm2}
\minimax(\mathbb R^p,N)\geq \frac{p^3}{N^2}\underbrace{\frac{4 |A_0|^3}{27}}_{c\approx 1.89}.
\end{equation}
where $A_0\approx -2.34$ is the first zero of the Airy function $\t{Ai}(\cdot)$.

\textbf{Theorem 3.} Assuming that $\Theta$ is not degenerated in any direction, i.e. there exists finite-size cube $[-\delta/2,+\delta/2]^p\in\Theta$,
the leading term $\sim \frac{1}{N^2}$ of $\minimax(\Theta,N)$ does not depend on $\Theta$:
\begin{equation}
\label{thm3a}
\forall_{\Theta}\lim_{N\to\infty} N^2\minimax(\Theta,N)=\lim_{N\to\infty} N^2\minimax(\mathbb R^p,N).
\end{equation}
Moreover, for finite $N$, satisfying $pN\delta\geq 2$, the following lower bound holds:
\begin{equation}
\label{thm3b}
\minimax(\Theta,N)\geq\minimax([-\delta/2,+\delta/2]^p,N)\geq \frac{cp^3}{N^2}\left(1-\frac{8p\log(pN\delta)}{N\delta}\right).
\end{equation}

\label{app:size}
\subsection{Irrelevance of adaptiveness in deriving the fundamental bound in estimation of completely uknown phases (proof of Theorem 1)}
\label{app:adaptive}
We start the proof by presenting an argument that in case of estimation of completely unknown phases, 
general adaptive estimation strategies do not provide any fundamental advantage compared with the strategy where all the photons are send through the interferometer together (provided they can be prepared in an arbitrary multimode entangled state) --- i.e. the strategy where all $V_i=\openone$.
This argument was shared with us  by Dominic Berry \cite{Dominic}.

First, let us discuss the structure of the output state \eqref{generalstate} (this is analogous to what was considered in \cite{Gorecki2020pi} in the case of single parameter estimation). As each action of the gate $U_\bvar$ multiplies its eigenvectors by the factor $e^{i\bmu\bvar}$: $U_{\bvar}=\int \t{d}\bmu\,e^{i\bmu \bvar}\ket{\bmu}\bra{\bmu}$, the  final state \eqref{generalstate} may be written as:
\begin{multline}
\ket{\Psi^N_\bvar}=V_N(U_\bvar\otimes \openone_A) V_{N-1}...V_1(U_\bvar\otimes \openone_A)\ket{\Psi_{\ins}}=\\
\int \t{d}\bmu_{(N)}...\int \t{d}\bmu_{(1)}e^{i(\bmu_{(N)}+...\bmu_{(1)})\bvar}\ket{\bmu_{(N)}}\left(\prod_{i=1}^{N-1}\braket{\bmu_{(i+1)}}{V_i|\bmu_{(i)}}\right)\braket{\bmu_{(1)}}{\Psi_{\ins}}=\\
\int \t{d}(\bmu_{(1)}+...\bmu_{(N)})e^{i (\bmu_{(1)}+...\bmu_{(N)})\bvar}
\int \t{d}\bmu^{\perp}\ket{\bmu_{(N)}}\left(\prod_{i=1}^{N-1}\braket{\bmu_{(i+1)}}{V_i|\bmu_{(i)}}\right)\braket{\bmu_{(1)}}{\Psi_{\ins}},
\end{multline}
where we have split the integral into integration over $(\bmu_{(1)}+...\bmu_{(N)})$ and $\bmu^{\perp}$ which represents all directions for which $(\bmu_{(1)}+...+\bmu_{(N)})$ is constant and also includes any necessary normalization factors. Note that in the above notation $\braket{\bmu_{(1)}}{\Psi_{\ins}}\in\mathcal H_A$, as well as $\braket{\bmu_{(i+1)}}{V_i|\bmu_{i}}$ is an operator acting on $\mathcal H_A$. Next, by rescaling the variable in the first integral $\bmu=\frac{\bmu_{(1)}+...\bmu_{(N)}}{N}$ and naming the last one by $\frac{1}{N}c(\bmu)\ket{g_\bmu}$ (where $\braket{g_{\bmu}}{g_{\bmu}}=1$) we end with:
\begin{equation}
\label{ada}
\ket{\Psi^N_\bvar}=\int\displaylimits_{\forall \mu_i\geq 0, \sum_i{\mu_i}\leq 1} \t{d}\bmu\, e^{iN\bmu\bvar}c(\bmu)\ket{g_\bmu}.
\end{equation}
Note, that $\ket{g_\bmu}$ constructed in such a way are not necessary mutually orthogonal, which in general makes the class of the states of the form \eqref{ada} larger than that in \eqref{simplystate}.
For a given measurement $\{M_{\tilde\bvar}\}$ the cost maximized over $\bvar\in\mathbb R^p$ is given by:
\begin{equation}
\max_{\bvar\in\mathbb R^p}\int \t{d}\tilde\bvar \bra{\Psi^N_\bvar}M_{\tilde\bvar}\ket{\Psi^N_\bvar}(\tilde\bvar-\bvar)^2.
\end{equation}

Next we show, that for any state of the form \eqref{ada} and the measurement $\{M_{\tilde\bvar}\}$ there exists a non-adaptive strategy with at least the same final precision. Consider the input state of the form $\propto\int c(\bmu)\ket{\bmu}d\bmu$.
After a free evolution, as a part of the measurement procedure, we entangle each $\ket{\bmu}$ with an ancillary system in $\ket{g_\bmu}$ state (this is possible since $\ket{\bmu}$ basis is orthogonal) obtaining:
\begin{equation}
\ket{\Psi^{N'}_\bvar} \propto \int c(\bmu)e^{iN\bmu\bvar}\ket{\bmu}\otimes\ket{g_\bmu}d\bmu
\end{equation}
and perform a measurement in the Fourier basis on the initial system:
\begin{equation}
\ket{\bavar}=\frac{1}{\sqrt{(2\pi/N)^p}}\int \t{d}\bmu\,e^{iN\bmu\bavar}\ket{\bmu}.
\end{equation}
The resulting state, given outcome $\bavar$, reads
\begin{equation}
\ket{\Psi^{N''}_\bvar(\bavar)} \propto \int d\bmu\, c(\bmu)e^{iN\bmu(\bvar-\bavar)}\ket{g_\bmu}=\ket{\Psi^{N}_{\bvar-\bavar}}
\end{equation}
and is obtained with probability distribution $p(\bavar)=\bra{\Psi^{N'}_{\bvar}}\ket{\bavar}\bra{\bavar}\otimes\openone\ket{\Psi^{N'}_{\bvar}}$. Next we apply the measurement which is optimal for the output state \eqref{ada}, and properly shift the indication of estimator by the 
measured value $\bavar$. In consequence, the mean cost for the true value of parameter $\bvar$ will be the same as the cost in the previously considered adaptive strategy \eqref{ada} for the value of parameter $\bvar-\bavar$, averaged with probability distribution $p(\bavar)$:
\begin{equation}
\max_{\bvar\in\mathbb R^p}\int \t{d}\tilde\bvar \int \t{d}\bavar\, p(\bavar)  \bra{\Psi^N_{\bvar-\bavar}} M_{\tilde\bvar-\bavar} \ket{\Psi^N_{\bvar-\bavar}}(\tilde\bvar-\bvar)^2.
\end{equation}
As averaging the cost may only \emph{decrease} the minimax bound, the statement is proven.

Therefore, from now on, when looking for the optimal strategy, we may focus on the output states of the form \eqref{simplystate}. Moreover, for technical reasons, we will treat the function $f(\bmu)$ as a function defined on the whole $\mathbb R^p$, but with the condition, that it is equal to $0$ outside of the region $\forall \mu_i\geq 0, \sum_i{\mu_i}\leq 1$.

Note that when only non-adaptive strategy is considered, it is a covariant problem of the group element estimation. Therefore, from the Hunt-Stein lemma~\cite{HOLEVO1979385,OZAWA198011} (generalized for noncompact group case~\cite{Ozawa1980,bogomolov1982minimax,hayashi2016fourier}) it follows that the search for the optimal measurement strategy may be restricted to projective covariant measurements \cite{Holevo1982}:
\begin{equation}
\ket{\chi_{\tilde\bvar}}=\frac{1}{\sqrt{(2\pi/N)^{p}}}\int_{\mathbb R^p}\t{d}\bmu\, e^{iN\bmu\tilde\bvar}\ket{\bmu}.
\end{equation}
One can also provide an intuitive argument: as the problem is exactly equivalent to the position-shift estimation (when $\bvar$ is treated as position), or momentum shift estimation problem (when $\bvar$ is treated as momentum), the optimal way is to measure this observable directly using measurement operators projecting on the observable eigenbasis~\cite{Tsang2020}.

After applying the covariant measurement we get:
\begin{multline}
\Delta^2\tilde\bvar \geq  \min_f \int\displaylimits_{\mathbb R^p} \t{d} \tilde{\bvar}\, |\braket{\chi_{\tilde{\bvar}}}{{\Psi_{f,\bvar}^N}}|^2 (\tilde\bvar-\bvar)^2  =\min_f \int\displaylimits_{\mathbb R^p} \t{d} \tilde{\bvar}\, |\braket{\chi_{\tilde{\bvar}}}{{\Psi_{f,0}^N}}|^2\tilde\bvar^2=
\min_f \int\displaylimits_{\mathbb R^p} \t{d} \tilde{\bvar}\,\left|\frac{1}{\sqrt{(2\pi/N)^{p}}}\int \t{d}\bmu e^{-iN\bmu\tilde\bvar}f(\bmu)
\right|^2\tilde\bvar^2
\\
= \min_{f}\int\displaylimits_{\mathbb R^p} \t{d} (N\tilde{\bvar})\,\left|\hat{f}(N\tilde\bvar)\right|^2\tilde\bvar^2 \overset{N\tilde\bvar\to\tilde\bvar}{=} \frac{1}{N^2}\min_{f}\int\displaylimits_{\mathbb R^p} \t{d} \tilde{\bvar}\,
\left|\hat{f}(\tilde\bvar)\right|^2\tilde\bvar^2,
\end{multline}
where in the first step we dropped the irrelevant dependence on $\bvar$, $\hat{f}$ is the Fourier transform of $f$ and at the end we rescaled $N\tilde\bvar\to\tilde\bvar$ to move the factor $\frac{1}{N^2}$ in front of the integral. Going back to the position representation, and recalling the constraints on $f$, we get \eqref{thm1}.

\subsection{Bound on the ground energy of a particle in a simplex potential well via a single degree of freedom formula  (proof of Theorem 2)}
\label{app:singlemodebound}
Given a general multimode pure state:
\begin{equation}
\ket{\Psi_{f}^N}=\int\displaylimits_{\forall \mu_i\geq 0, \sum_i{\mu_i}\leq 1}d\boldsymbol{\mu}\, f(\boldsymbol{\mu}) \ket{\mu_1,\mu_2,..\mu_p}
\end{equation}
let us define a single mode reduced density matrix corresponding to the $i$-th sensing arm mode:
\begin{equation}
\label{densitym}
\rho_i(\mu_i;\mu_i^\prime) = \int \left(\prod_{j \neq i}\t{d}\mu_j \t{d}\mu_j'\, \delta(\mu_j-\mu_j')\, \right) f(\boldsymbol{\mu}) f^*(\boldsymbol{\mu}').
\end{equation}
The formula for the lower bound on the estimation cost \eqref{thm1} may be equivalently written in terms of the single mode reduced density matrices:
\begin{equation}
\Delta^2\tilde\bvar \geq \frac{1}{N^2}\min_{f(\boldsymbol{\mu})}\int\displaylimits_{\forall \mu_i\geq 0, \sum_i{\mu_i}\leq 1}
\sum_{i=1}^p \t{d}\mu_i \t{d} \mu_i' \, \delta(\mu_i-\mu_i')\left(-\frac{\partial^2}{\partial\mu_i^2}\right)\ \rho_i(\mu_i;\mu_i').
\end{equation}
Due to the inherent symmetry of the problem, we assume all the reduced density matrices to be identical and hence we do not distinguish them by a subscript $i$. As discussed in the main text, this symmetry implies also that the expectation value of the number of photons in each single `phase arm' is $\leq N/p$. Therefore, looking for the lower bound of the above formula, we may relax the original constraints by ignoring the fact that $\rho(\mu;\mu)$ comes from the multimode state satisfying conditions from \eqref{thm1}, and perform optimization with a weaker single mode constraints:
\begin{equation}
\quad \rho(0;0) = 0, \quad \int \t{d}\mu \, \rho(\mu;\mu)  = 1, \quad \int \t{d}\mu \, \rho(\mu;\mu) N\mu \leq \sfrac{N}{p}, \quad \rho \geq 0.
\end{equation}
This allows us to write the lower bound as:
\begin{equation}
\label{densmatrixbound}
\Delta^2\tilde\bvar \geq p\times\frac{1}{N^2} \min_{\rho}
\int_0^\infty \t{d}\mu \t{d} \mu' \, \delta(\mu-\mu')\left(-\frac{\partial^2}{\partial\mu^2}\right)\ \rho(\mu;\mu')
\end{equation}
with the above constraints imposed. 

Let $\bar{\rho}(\mu;\mu')$ be the solution of the above minimization problem.
It implies that $\int \t{d}\mu \, \bar{\rho}(\mu;\mu) N\mu = \sfrac{N}{p}$ 
(note that we have replaced the inequality with the equality here, since any function with average photon number $t\cdot \sfrac{N}{p}$, where $0<t<1$, can always be rescaled $\bar{\rho}(\mu;\mu')\to t\bar{\rho}(t\mu;t\mu')$, which reduces the cost without breaking constrains). If, in addition, the solution corresponds to a pure state
$\bar{\rho}(\mu;\mu') = g(\mu) g^*(\mu')$ then we indeed arrive at formula for the bound  \eqref{eq:singlemodebound}
stated in the main text.

\begin{equation}
\Delta^2\tilde\bvar \geq p\times\frac{1}{N^2}
\min_g \int_0^\infty \t{d}\mu\, g^*(\mu)\left(-\frac{\partial^2}{\partial\mu^2}\right)g(\mu)
\end{equation}
with constraints:
\begin{equation}
\label{cons}
 g(0)=0,\ \int_0^\infty \t{d}\mu\, g^*(\mu)g(\mu)=1,\  \int_0^\infty \t{d}\mu\, g^*(\mu)g(\mu)N\mu =\sfrac{N}{p}.
\end{equation}
The solution may be found using the standard Lagrange multiplier method,
\begin{equation}
-\frac{\partial^2}{\partial \mu^2}g(\mu)+g(\mu)(\lambda_1+\mu\lambda_2)=0 \Rightarrow g(\mu)\propto \t{Ai}\left(\frac{\lambda_1+\lambda_2\mu}{\lambda_2^{2/3}}\right),
\end{equation}
where $\t{Ai}(\cdot)$ is the Airy function of the first kind and after taking into account conditions \eqref{cons} we obtain:
\begin{equation}
\label{eq:g}
g(\mu)= \frac{1}{\t{Ai}'(A_0)}\sqrt{\frac{2p|A_0|}{3}}\t{Ai}\left(A_0+\frac{2p|A_0|}{3}\mu\right),
\end{equation}
where $A_0\approx -2.34$ is the first zero of the Airy function and $\t{Ai}'(\cdot)$ is its first derivative. 
The corresponding bound on the cost reads:
 \begin{equation}
\Delta^2\tilde\bvar\geq p\times\frac{p^2}{N^2}\frac{4 |A_0|^3}{27}=\frac{p^3}{N^2}\underbrace{\frac{4 |A_0|^3}{27}}_{c\approx 1.89}.
\end{equation}

What remains to be shown is that the assumption of a pure state is justified and that a mixed state cannot provide a lower value for the bound.
The solution of the pure state case implies that for the mean photon number $\bar{N}$ fixed (in this case $\bar{N} = N/p$), the resulting optimal cost (for single phase) is equal $\frac{1}{N^2}
\min_g \int_0^\infty \t{d}\mu\, g^*(\mu)\left(-\frac{\partial^2}{\partial\mu^2}\right)g(\mu)=c/\bar{N}^2$.
Consider now, instead of a pure state $\bar{\rho} = \ket{g}\bra{g}$, an exemplary mixed state which for concreteness we choose to be a probabilistic mixture of two pure states $\bar{\rho}' = p_1 \ket{g_1}\bra{g_1} + p_2 \ket{g_2} \bra{g_2}$.
The constraint on the mean photon number implies that
\begin{equation}
p_1 \bar{N}_1 + p_2 \bar{N}_2 = \bar{N},
\end{equation}
where $\bar{N}_k$ are mean photon numbers in states $\ket{g_k}$ respectively.
The corresponding cost for the mixed state will be a weighted sum for the costs of the respective pure states,
\begin{equation}
p_1 \frac{c}{\bar{N}_1^2} + p_2 \frac{c}{\bar{N}_2^2} \geq  \frac{c}{(p_1 \bar{N}_1 + p_2\bar{N}_2)^2} = \frac{c}{\bar{N}^2},
\end{equation}
where the inequality follows from the convexity property of the $1/x^2$ function (Jensen inequality).
This reasoning can be trivially generalized to mixed states involving mixtures of more than two states and proves that a mixed state cannot provide a lower value for the bound \eqref{densmatrixbound} than the optimal pure state.

\onecolumngrid
\subsection{Application for the local estimation problem  (proof of Theorem 3)}
We assume that the prior parameter region $\Theta$ has a finite volume and hence we can always find $\delta > 0$ such that
$[-\delta/2,+\delta/2]^p\subset\Theta$. Replacing $\Theta$ with its subset may only lower the minimax cost and hence any lower bound we obtain for the easier problem will also be valid for the original one. Therefore, in what follows we assume $\Theta = [-\delta/2,+\delta/2]^p$. 

For the purposes of the proof, let us imagine for the moment that instead of knowing that $\bvar\in[-\delta/2,+\delta/2]^p$,
we assume that $\bvar$ is completely unknown and hence the region to consider is $\bvar\in\mathbb R^p$.
Let us also assume that we have  $N+p\cdot N_0$ gates at our disposal and that we spend $p \cdot N_0$ resources to find an approximated values of the parameters $\tilde\bvar$, such that the true value of the $\bvar$ (with high probability) lies in its close neighborhood $[\tilde\var_1-\delta/2,\tilde\var_1+\delta/2]\times...\times[\tilde\var_p-\delta/2,\tilde\var_p+\delta/2] $. Next we use the remaining $N$ in the optimal way to estimate $\bvar$, taking into account this knowledge. This approach may be suboptimal, from the point o view of the optimal use of  all $N+p\cdot N_0$ resources, and hence we may write the following inequality:
\begin{equation}
\minimax(\mathbb R^p,N+p\cdot N_0)\leq \minimax([-\delta/2,+\delta/2]^p,N)+R(p,\delta,N_0),
\end{equation}
where in $\minimax([-\delta/2,+\delta/2]^p,N)$ we have shifted the region to the origin (which does not change the minimax cost). Here the term $R(p,\delta,N_0)$ corresponds to the risk, that true value of $\bvar$ lies outside of predicted region $[-\delta/2,+\delta/2]^p$ (formal definition will be given later). Intuitively, as the discrimination error decreases exponentially with $N_0$, for large value of $N$ it is enough to take $N_0$ sublinear in $N$ to make the term $R(p,\delta,N_0)$ small enough to become irrelevant. Therefore, we expect that the minimax costs for the  finite and infinite regions approach the same limit in the leading order. Below we discuss the rate of convergence.

\subsubsection{From no prior knowledge to finite knowledge}
Consider the following strategy. First, for each phase we use $N_0$ gates to get the state $\ket{\Psi_{\var_i}^{N_0}}=\int_0^{1}\t{d}\mu_i\, e^{iN_0\mu_i\var_i}w(\mu_i)\ket{\mu_i}$ (where all remaining $\mu_{j\neq i}=0$). After performing the covariant measurement we obtain the result $\tilde\var_i$ with probability:
\begin{equation}
\label{apriori}
|\braket{\chi_{\tilde\var_i}}{\Psi_{\var_i}^{N_0}}|^2=|\hat w(\var_i-\tilde\var_i)|^2=p_{N_0}(\underbrace{\var_i-\tilde\var_i}_{\dvar_i}).
\end{equation}
We repeat it for all phases an get $p_{N_0}(\bdvar)=\prod_{i=1}^p p_{N_0}(\dvar_i)$. In the next step we need to estimate the value of the difference $\bdvar=\bvar-\tilde\bvar$. Note that while $\bvar$ has a fixed value, $\bdvar$ is a random variable and hence we may make use of the Bayesian cost formulas. Since such a procedure of estimating $\bvar$ using $N+p \cdot N_0$ gates is in general suboptimal, the final mean variance may be bounded by the minimal cost when the total amount of resources $N+p \cdot N_0$ are used:
\begin{equation}
\label{old}
\minimax(\mathbb R^p,N+p\cdot N_0)\leq
\underbrace{\min_{M_{\tilde\bdvar},\rho_{\bdvar}^{N}}\int \t{d}\bdvar\,p_{N_0}(\bdvar)\int d\tilde\bdvar \tr(M_{\tilde\bdvar}\rho_{\bdvar}^{N})(\tilde\bdvar-\bdvar)^2}_{\bayes(p_{N_0},N)}.
\end{equation}
This bounds from below the minimal obtainable Bayesian cost with a priori distribution $p_{N_0}(\bdvar)$ and use of $N$ gates, $\bayes(p_{N_0},N)$.


Now we would like to use this result to state the bound for the minimax cost for $\bvar\in[-\delta/2,+\delta/2]^p$. By definition the minimax cost is greater or equal than the minimal Bayesian cost for any prior with support inside of $[-\delta/2,+\delta/2]^p$ (formal justification will be given in \eqref{mmvsbay}). As $p_{N_0}(\bdvar)$ is given by the square of modulus of the Fourier transform of a finite support function $w(\mu_i)$, $p_{N_0}(\bdvar)$ will always stick outside any finite size region $[-\delta/2,+\delta/2]^p$. However, in principle these tails may be arbitrary small if sufficiently large $N_0$ is allowed (the exemplary choice of $w(\mu_i)$ will be given in next section).

For the purpose of the proof, let us define the distribution corresponding to $p_{N_0}(\bdvar)$, but with the tails cut:
\begin{equation}
p_{N_0}^{\delta}(\dvar_i)=\begin{cases}
\frac{1}{1-R_1}p_{N_0}(\dvar_i)&{\text{ for }}\dvar_i\in[-\delta/2,+\delta/2]\\
0 &{\text{ for }}\dvar_i\notin[-\delta/2,+\delta/2]
\end{cases},
\end{equation}
where $\frac{1}{1-R_1}$ is a proper normalization factor, i.e.
\begin{equation}
\label{R1}
R_1=2\int_{\delta/2}^{+\infty}\t{d}\dvar_i\, p_{N_0}(\dvar_i).
\end{equation}

For such a distribution we may formally bound the minimax cost using the minimal Bayesian cost. Let $\rho_{\bdvar}^{N,{\rm minimax}}$, $M_{\tilde\bdvar}^{\rm minimax}$ be the output state and the measurement optimizing the minimax cost. Then:
\begin{multline}
\label{mmvsbay}
\minimax([-\delta/2,+\delta/2]^p,N)=\max_{\bdvar\in[-\delta/2,+\delta/2]^p} \int \t{d}\tilde\bdvar \tr(\rho_{\bdvar}^{N,{\rm minimax}}M_{\tilde\bdvar}^{\rm minimax})(\tilde\bdvar-\bdvar)^2\\
\geq \int_{[-\delta/2,+\delta/2]^p}p_{N_0}^{\delta}(\bdvar)\t{d}\bdvar \int \t{d}\tilde\bdvar \tr(\rho_{\bdvar}^{N,{\rm minimax}}M_{\tilde\bdvar}^{\rm minimax})(\tilde\bdvar-\bdvar)^2\\
\geq \min_{M_{\tilde\bdvar},\rho_{\bdvar}^{N}}\int_{[-\delta/2,+\delta/2]^p}p_{N_0}^{\delta}(\bdvar)\t{d}\bdvar \int \t{d}\tilde\bdvar \tr(\rho_{\bdvar}^{N}M_{\tilde\bdvar})(\tilde\bdvar-\bdvar)^2=\bayes(p_{N_0},N).
\end{multline}
Alternatively, the above inequality may be derived directly using the fact, that the optimal minimax cost is equal to the minimal Bayesian cost with the least favorable prior~\cite{tanaka2014quantum}.

Summarizing \eqref{old} and \eqref{mmvsbay} we have:
\begin{equation}
\label{idea}
\begin{split}
\minimax(\mathbb R^p,N+p\cdot N_0)&\leq \bayes(p_{N_0},N),\\
\bayes(p_{N_0}^\delta,N)&\leq \minimax([-\delta/2,+\delta/2]^p,N).
\end{split}
\end{equation}
What remains is to connect $\bayes(p_{N_0},N)$ with $\bayes(p_{N_0}^\delta,N)$.

Let $\rho_{\bdvar}^{\delta,N},M_{\tilde\bdvar}^{\delta}$ be the output state and the measurement minimizing the Bayesian cost for $p_{N_0}^{\delta}(\bdvar)$. Obviously $\tilde\bdvar$ that appear in $M_{\tilde\bdvar}^{\delta}$ are only inside of $[-\delta/2,+\delta/2]^p$. By applying this state and measurement with the original prior distribution $p_{N_0}(\bdvar)$ we get:
\begin{equation}
\label{cut0}
\bayes(p_{N_0},N)\leq\int_{\mathbb R^p} \t{d}\bdvar\, p_{N_0}(\bdvar)\int_{[-\delta/2,+\delta/2]^p} \t{d}\tilde\bdvar\,\tr(M_{\tilde\bdvar}^{\delta}\rho_\bdvar^{\delta,N})(\tilde\bdvar-\bdvar)^2.
\end{equation}
Expanding $(\tilde\bdvar-\bdvar)^2 = \sum_i (\tilde{\dvar}_i-\dvar_i)^2$, and considering a single term from the sum corresponding to $\dvar_i$,
we can divide the resulting integrals appearing on the RHS of the above equation into three parts:
\begin{equation}
\begin{split}
\label{cut}
&\int_{\mathbb R^p}\t{d}\bdvar\, p_{N_0}(\bdvar)\int_{[-\delta/2,+\delta/2]} \t{d}\tilde\dvar_i\, \tr\left[\underbrace{\left(\int_{[-\delta/2,+\delta/2]^{p-1}}\t{d}\tilde\dvar_{j\neq i}\,M_{\tilde\bdvar}^{\delta}\right)}_{\overline{M}^\delta_{\dvar_i}}\rho_\bdvar^{\delta,N}\right](\tilde\dvar_i-\dvar_i)^2=\\
(a)\quad&\int_{[-\delta/2,+\delta/2]^p}\t{d}\bdvar\,  p_{N_0}(\bdvar)\int_{[-\delta/2,+\delta/2]}\t{d}\tilde\dvar_i\, \tr\left[
\overline{M}^\delta_{\dvar_i}
\rho_\bdvar^{\delta,N}\right](\tilde\dvar_i-\dvar_i)^2+\\
(b)\quad&\int_{[-\delta/2,+\delta/2]} \t{d}\dvar_i\, p_{N_0}(\dvar_i)
\int_{[-\delta/2,+\delta/2]}\t{d}\tilde\dvar_i\, \tr\left[
\overline{M}^\delta_{\dvar_i}
\underbrace{\int_{\mathbb R^{p-1}\setminus [-\delta/2,+\delta/2]^{p-1}}\t{d}\dvar_{j\neq i}\,p(\dvar_j)\rho_\bdvar^{\delta,N}}_{\overline \rho^{\delta,N}_{\dvar_i}}\right](\tilde\dvar_i-\dvar_i)^2+\\
(c)\quad&\int_{\mathbb R\setminus [-\delta/2,+\delta/2]}\t{d}\dvar_i\, p_{N_0}(\dvar_i)
\int_{[-\delta/2,+\delta/2]}\t{d}\tilde\dvar_i\, \tr\left[
\overline{M}^\delta_{\dvar_i}
\int_{\mathbb R^{p-1}}\t{d}\dvar_{j\neq i}\,p(\dvar_j)\rho_\bdvar^{\delta,N}\right](\tilde\dvar_i-\dvar_i)^2.
\end{split}
\end{equation}
Part $(a)$ represents the contribution  where $\bdvar \in [-\delta/2,+\delta/2]^p$, so after summing over all $i$ it is exactly equal $(1-R_1)^p\cdot\bayes(p_{N_0}^\delta,N)$.
Part $(b)$ corresponds to the case where $\dvar_i$ belongs to $[-\delta/2,+\delta/2]$ (probability $1-R_1$) but at least one of $\dvar_{j\neq i}$ does not (probability $1-(1-R_1)^{(p-1)} = \tr(\overline{\rho}^{\delta,N}_{\dvar_i})$). It means that this integral may be bounded from above by the worst possible estimation strategy (resulting in squared error $\delta^2$) multiplied by this probability---$(1-R_1)(1-(1-R_1)^{(p-1)})\delta^2$. For our purpose it is enough to use a weaker, but simpler, bound $(1-R_1)(1-(1-R_1)^{(p-1)})\delta^2\leq (p-1)R_1 \delta^2$.

Finally $(c)$, where $\dvar_i$ is outside of $[-\delta/2,+\delta/2]$, may be bounded from above by:
\begin{equation}
\label{R2}
R_2=2\int_{\delta/2}^{+\infty}\t{d}\dvar_i\, p_{N_0}(\dvar_i)(\dvar_i-(-\delta/2))^2,
\end{equation}
where the most pessimistic scenario is assumed (each time when the true value of $\dvar_i$ is to the right of $[-\delta/2,+\delta/2]$, our estimator point is at the left border of the region $-\delta/2$; similarly in the opposite situation).

Combining all that  was said above and taking the sum over $p$ parameters, we may bound the RHS of \eqref{cut0} by:
\begin{equation}
(1-R_1)^p\bayes(p_{N_0}^\delta,N)+\underbrace{p(p-1) R_1\delta^2+p R_2}_{R(p,\delta,N_0)},
\end{equation}
and we arrive at:
\begin{equation}
\bayes(p_{N_0},N)\leq (1-R_1)^p\bayes(p_{N_0}^\delta,N)+R(p,\delta,N_0).
\end{equation}
By rearranging this inequality and using $(1-R_1)<1$ we get:
\begin{equation}
\label{singlebound}
\bayes(p_{N_0}^\delta,N)\geq\bayes(p_{N_0},N)-R(p,\delta,N_0).
\end{equation}
After applying the above to \eqref{idea} we end up with:
\begin{equation}
\label{eq:unknownfinite}
\minimax([-\delta/2,+\delta/2]^p,N)\geq\minimax(\mathbb R^p,N+p\cdot N_0)- R(p,\delta,N_0),
\end{equation}
which was to be proved. What remains is to find $w(\mu_i)$ (with corresponding $p_{N_0}(\dvar_i)$), for which $R(p,\delta,N_0)$ decrease sufficiently fast with $N_0$.

\subsubsection{Irrelevance ot the size of $\Theta$ in the limit $N\to\infty$}
\label{app:integrals}
Following \cite{Gorecki2020pi} we choose $w(\mu_i)$ to be proportional to the self-convolution of the Kaiser window function with window duration $N_0/2$. Then $p_{N_0}(\dvar_i)$ is  proportional to the fourth power of the Fourier transform of the Kaiser window function (from now on in this section the subscript $i$ in $\dvar_i$ will be omitted to simplify notation):
\begin{equation}
p_{\alpha,L}(\dvar)=\mathcal N_\alpha L {\rm sinc}^4\left(\pi\alpha\sqrt{(L\dvar/4\alpha)^2-1}\right)=\mathcal N_\alpha L\frac{{\rm sinh}^4\left(\pi\alpha\sqrt{1-(L\dvar/4\alpha)^2}\right)}{\left(\pi\alpha\sqrt{1-(L\dvar/4\alpha)^2}\right)^4},
\end{equation}
where $L=2N_0$ is the bandwidth, $\alpha$ determines the shape and $\mathcal N_\alpha$ may be bounded by:
\begin{equation}
\label{Nbound}
\mathcal N_\alpha\lesssim4\sqrt{2}\pi^4\alpha^{7/2}e^{-4\pi\alpha},
\end{equation}
where the bound is tight for big $\alpha$~\cite{Gorecki2020pi}. As shown in \cite{Gorecki2020pi} only exponentially small (with $\alpha$) part lays outside of region $[-4\alpha/L,4\alpha/L]$ and, therefore, for our purpose we choose $\delta/2=4\alpha/L$.
We may bound $R_1$ and $R_2$ by:
\begin{multline}
\label{R1b}
R_1= 2\mathcal N_{\alpha} \int^{+\infty}_{4\alpha/L} \t{d}\dvar\, L {\rm sinc}^4\left(\pi\alpha\sqrt{(L\dvar/4\alpha)^2-1}\right)
=2\mathcal N_{\alpha} \frac{4\alpha}{L}\int^{+\infty}_{1}\t{d}x L {\rm sinc}^4\left(\pi\alpha\sqrt{x^2-1}\right) \leq\\
\leq  8\mathcal N_{\alpha} \alpha\Big(\int_1^2 \t{d}x+\frac{1}{\pi^4\alpha^4}\int_2^\infty \t{d}x\frac{1}{(x^2-1)^2}\Big)
=8\mathcal N_{\alpha} \alpha\Big(1+\frac{1/3-\log(3)/4}{\pi^4\alpha^4}\Big)\overset{\alpha>1/2}{\leq}16\mathcal N_\alpha \alpha,
\end{multline}


\begin{multline}
\label{R2b}
R_2= 2\mathcal N_{\alpha} \int^{+\infty}_{4\alpha/L} \t{d}\dvar\, L {\rm sinc}^4\left(\pi\alpha\sqrt{(L\dvar/4\alpha)^2-1}\right)(\var+\delta/2)^2
=2\mathcal N_{\alpha} \frac{4\alpha}{L}\int^{+\infty}_{1}\t{d}x\, L {\rm sinc}^4\left(\pi\alpha\sqrt{x^2-1}\right)(x+1)^2\left(\frac{4\alpha}{L}\right)^2\leq\\
\leq 2\mathcal N_{\alpha} L\left(\frac{4\alpha}{L}\right)^3\Big( \int_1^2\t{d}x\, (x+1)^2+\frac{1}{\pi^4\alpha^4}\int_2^{+\infty}\t{d}x\, \frac{(x+1)^2}{(x^2-1)^2}\Big)
=2\mathcal N_{\alpha} L\left(\frac{4\alpha}{L}\right)^3\Big(\frac{19}{3}+\frac{1}{\pi^4\alpha^4}\Big)\overset{\alpha>1/2}{\leq}14\mathcal N_\alpha L\left(\frac{4\alpha}{L}\right)^3,
\end{multline}
where inequalities ${\rm sinc}(x)\leq 1$ and ${\rm sinc}(x)\leq 1/x$ were used. We have:
\begin{multline}
R(p,\delta, N_0)=p(p-1)R_1\delta^2+pR_2\leq16 p^2 \mathcal N_{\alpha} \alpha\left(\frac{8\alpha}{L}\right)^2+14 p \mathcal N_{\alpha} L\left(\frac{4\alpha}{L}\right)^3=
\mathcal N_{\alpha}\frac{\alpha^3}{L^2}(1024p^2+896p)\leq 1920\mathcal N_{\alpha}\frac{\alpha^3}{L^2}p^2,
\end{multline}
which after substituting $L=2N_0$, $\alpha=N_0\delta/4$, yields:
\begin{equation}
\label{Rbound}
R(p,\delta,N_0)\leq \frac{15}{2}\mathcal N_{N_0\delta/4}N_0\delta^3 p^2.
\end{equation}
To derive \eqref{thm3a}, we use a trivial fact $\forall_{\Theta}\minimax(\mathbb R^p,N)\geq\minimax(\Theta,N)$ and \eqref{eq:unknownfinite}
and we get:
\begin{equation}
\minimax(\mathbb R^p,N)\geq \minimax(\Theta,N)\geq \minimax(\mathbb R^p,N+p\cdot N_0)- \frac{15}{2}\mathcal N_{N_0\delta/4}N_0\delta^3 p^2.
\end{equation}
Next, as $\minimax(\mathbb R^p,N+p\cdot N_0)=\frac{N^2}{(N+p N_0)^2}\minimax(\mathbb R^p,N)$, we choose $N_0=\sqrt{N}$ and apply $\lim_{N\to\infty}N^2\cdot$ to the above inequality in order to get:
\begin{equation}
\lim_{N\to\infty}N^2\minimax(\mathbb R^p,N)\geq \lim_{N\to\infty}N^2\minimax(\Theta,N)\geq \lim_{N\to\infty}N^2\minimax(\mathbb R^p,N),
\end{equation}
which was to be proven. In order to prove \eqref{thm3b} more subtle choice of $N_0$ will be required.

\subsubsection{Convergence of the bound for finite-size $\Theta$}

From \eqref{thm2} we have
\begin{equation}
\minimax(\mathbb R^p,N+p\cdot N_0)\geq\frac{cp^3}{(N+p\cdot N_0)^2}=\frac{cp^3}{N^2}\left(1-\frac{2pN_0}{N}+\frac{3(pN_0/N)^2+2(pN_0/N)^3}{(1+pN_0/N)^2}\right),
\end{equation}
with $c=\frac{4|A_0|^3}{27}\approx 1.89$. Therefore,
when combining this formula with \eqref{eq:unknownfinite}, we obtain a corresponding bound
for the finite prior case:
\begin{equation}
\label{ef}
\minimax([-\delta/2,+\delta/2]^p,N)\geq\frac{cp^3}{N^2}\left(1-\frac{2pN_0}{N}+\frac{3(pN_0/N)^2+2(pN_0/N)^3}{(1+pN_0/N)^2}-\frac{N^2R(p,\delta,N_0)}{cp^3}\right).
\end{equation}
To make $R(p,\delta,N_0)$ sufficiently small regardless of $p$ and $\delta$, we choose $N_0=\frac{4}{\delta}\log(p N\delta)$.
Next we will show that the difference of two last terms appearing in above equation is strictly positive for sufficiently big $pN\delta$, so they may be neglected without breaking inequality. To do that, let us bound it from below by the function of $pN\delta$.

Note that $\frac{3x^2+2x^3}{(1+x)^2}$ is monotonically increasing for positive $x$ and therefore:
\begin{equation}
\frac{3(pN_0/N)^2+2(pN_0/N)^3}{(1+pN_0/N)^2}\geq \frac{3(N_0/(pN))^2+2(N_0/(pN))^3}{(1+N_0/(pN))^2}=\frac{3(4\log(pN\delta)/(pN\delta))^2+2(4\log(pN\delta)/(pN\delta))^3}{(1+4\log(pN\delta)/(pN\delta))^2},
\end{equation}
where in the first step $pN_0/N$ was replaced by $N_0/(pN)$ everywhere. Moreover, using \eqref{Rbound}:
\begin{equation}
\frac{N^2R(p,\delta,N_0)}{cp^3}\leq \frac{N^2}{cp^3}\frac{15}{2}\mathcal N_{N_0\delta/4}N_0\delta^3p^2\leq \frac{N^2}{c}\frac{15}{2}\mathcal N_{N_0\delta/4}N_0\delta^3p^2=\frac{30}{c}(pN\delta)^2\mathcal N_{\log(pN\delta)}\log(pN\delta),
\end{equation}
where in the second step the factor $p^3$ was removed from denominator.

Taking two above together, naming by $y=pN\delta$ an using the bound for $\mathcal N_{\log(pN\delta)}$ \eqref{Nbound} we have therefore: 
\begin{equation}
\label{boundR}
\frac{3(pN_0/N)^2+2(pN_0/N)^3}{(1+pN_0/N)^2}-\frac{N^2R(p,\delta,N_0)}{cp^3}\geq \frac{48(\log(y)/y)^2+128(\log(y)/y)^3}{(1+4\log(y)/y))^2}-\frac{120\sqrt{2}\pi^4}{c}y^{2-4\pi}\log^{9/2}(y)
\end{equation}
It may be checked numerically, that for $c\geq 1.89$ it is strictly positive for $y\geq 2$. Therefore, for $pN\delta\geq 2$ (which justify also usage $\alpha>1/2$ in \eqref{R1b} and \eqref{R2b}), \eqref{ef} implies:
\begin{equation}
\label{bc}
\minimax([-\delta/2,+\delta/2]^p,N)\geq \frac{cp^3}{N^2}\left(1-\frac{8p\log(pN\delta)}{N\delta}\right).
\end{equation}



\subsubsection{Relation with the single-parameter Bayesian estimation bound}
For completeness, we show here how the above reasoning might be used to rederive the Bayesian bound for the single parameter unitary estimation with finite bandwidth prior, Eq. (12) from \cite{Gorecki2020pi}, as well as the minimax analogy of the Bayesian bound for rectangular prior of width $\delta$,  Eq. (14) from \cite{Gorecki2020pi}. Note the difference in notation as the $N$ from this paper corresponds to $n$ from \cite{Gorecki2020pi}.

Indeed, starting with \eqref{old}, and using the fact that the minimal achievable cost, when estimating a completely unknown single phase using $N$ phase gates, is given by  $\pi^2/N^2$, we get:
\begin{equation}
\bayes(p_{N_0},N)\geq \frac{\pi^2}{(N+N_0)^2}.
\end{equation}
Next, we extend the result to an arbitrary generator with spectrum in $[\lambda_-,\lambda_+]$ (instead of $\{0,1\}$), which corresponds to the replacement $N\to N(\lambda_+-\lambda_-), N_0\to N_0(\lambda_+-\lambda_-)$ in RHS of the equation above. Finally, from \eqref{apriori}, as $\hat w(\var)$ has a finite banwidth of size $N_0(\lambda_+-\lambda_-)$, the bandwidth of $p_{N_0}(\var)$ is at most $2N_0(\lambda_+-\lambda_-)$. Moreover, from \cite{Gorecki2020pi,boas1995}, for any non-negative integrable finite bandwidth function $p_{L}(\var)$ there exists a proper function $\hat w(\var)$ with bandwidth $L/2$ satisfying $p_{L}(\var)=|\hat w(\var)|^2$. Therefore:
\begin{equation}
\bayes(p_{L},N)\geq\frac{\pi^2}{(N(\lambda_+-\lambda_-)+L/2)^2},
\end{equation}
which is exactly Eq. (12) from \cite{Gorecki2020pi}.

Analogously, using \eqref{bc} with $c\to\pi^2$, $p\to1$,
$N\to N(\lambda_+-\lambda_-)$ we get:

\begin{equation}
\minimax([-\delta/2,+\delta/2],N)\geq
 \frac{\pi^2}{[N(\lambda_+-\lambda_-)]^2}\left(1-\frac{8\log(N(\lambda_+-\lambda_-)\delta)}{N(\lambda_+-\lambda_-)\delta}\right),
\end{equation}
which is similar to the bound for the Bayesian cost Eq. (14) from \cite{Gorecki2020pi}, but converges faster due to the lack of square root over the last term.

\section{Demonstrating joint phase estimation advantage over separate strategies using the state given in \eqref{eq:state}}
\label{app:B}
\label{app:state}
\subsection{Joint phase estimation cost}
For a given (not normalized) function:
\begin{equation}
\label{appstate}
f(\bmu)=\left(\prod_{i=1}^p\mu_i\right)^\alpha\left(1-\sum_{i=1}^p\mu_i\right)^\beta,
\end{equation}
(where $\alpha,\beta\geq \frac{1}{2}$) we will first calculate the normalization factor:
\begin{equation}
\mathcal N=\int\displaylimits_{\forall \mu_i\geq 0, \sum_i{\mu_i}\leq 1}d\boldsymbol{\mu} f^*(\boldsymbol{\mu})f(\boldsymbol{\mu})
\end{equation}
and then the corresponding `mean energy':
\begin{equation}
\mathcal E=\int\displaylimits_{\forall \mu_i\geq 0, \sum_i{\mu_i}\leq 1}d\boldsymbol{\mu} f^*(\boldsymbol{\mu})\left(\sum_{i=1}^{p}-\partial_i^2\right)f(\boldsymbol{\mu}).
\end{equation}
The final estimation cost will correspond to the ratio $1/N^2\cdot \mathcal E/\mathcal{N}$.

Let us introduce the objects representing `(P)roduct' and '$1$ minus (S)um' of coordinates:
\begin{equation}
P_{k}=\prod_{i=1+k}^p \mu_i,\quad S_{k}=1- \sum_{i=1+k}^p\mu_i,
\end{equation}
so $k$ indicates which coordinates are dropped. Then $f(\boldsymbol{\mu})=P_0^\alpha S_0^\beta$ and:
\begin{equation}
\mathcal N=\int_{0}^{S_p}\t{d}\mu_p \cdots  \int_{0}^{S_2}\t{d}\mu_2
\int_0^{S_{1}}\t{d}\mu_1 P_{0}^{2\alpha} S_{0}^{2\beta}.
\end{equation}
For the first integral we have:
\begin{equation}
\int_0^{S_{1}} \t{d}\mu_{1}\, P_{0}^{2\alpha} S_{0}^{2\beta}=\int_0^{S_{1}}  (P_{1}\mu_1)^{2\alpha} (S_{1}-\mu_1)^{2\beta}\t{d}\mu_{1}
=P_1^{2\alpha}S_1^{2\beta+(1+2\alpha)}\frac{\Gamma(1+2\alpha)\Gamma(1+2\beta)}{\Gamma(1+2\beta+(1+2\alpha))}.
\end{equation}
For the $k^{\t{th}}$ one:
\begin{equation}
\int_0^{S_{k}} \t{d}\mu_{k}\, P_{k-1}^{2\alpha} S_{k-1}^{2\beta+(k-1)(1+2\alpha)}=
\int_0^{S_{k}} \t{d}\mu_{k}\,  (P_{k}\mu_k)^{2\alpha} (S_{k}-\mu_k)^{2\beta+(k-1)(1+2\alpha)}
=P_k^{2\alpha}S_k^{2\beta+k(1+2\alpha)}C_{\alpha,\beta,k},
\end{equation}
where
\begin{equation}
C_{\alpha,\beta,k}=\frac{\Gamma(1+2\alpha)\Gamma(1+2\beta+(k-1)(1+2\alpha))}{\Gamma(1+2\beta+k(1+2\alpha))}.
\end{equation}
As $P_p=1$, $S_p=1$, the $\mathcal N$ is therefore given by:
\begin{equation}
\mathcal N=\prod_{k=1}^p C_{\alpha,\beta,k}=\frac{\Gamma(1+2\alpha)^p\Gamma(1+2\beta)}{\Gamma(1+2\beta+p(1+2\alpha))}.
\end{equation}
Due to symmetry of $f(\boldsymbol{\mu})$, the mean value of $-\partial_i^2$ operator is the same for each coordinate. Therefore it is sufficient to calculate it for $i=1$. We have
\begin{equation}
|\partial_1 f(\bmu)|^2=\left(\partial_1(P_1\mu_1)^\alpha(S_1-\mu_1)^\beta\right)^2
=(S_1-\mu_1)^{2(\beta-1)}(P_1\mu_1)^{2\alpha}(\alpha S_1-(\alpha+\beta)\mu_1)^2\mu_1^{-2},
\end{equation}
so for the first integral we have:
\begin{equation}
\int_0^{S_{1}} \t{d}\mu_1\, |\partial_1 f(\bmu)|^2=
P_1^{2\alpha}S_1^{-1+2\alpha+2\beta}\frac{\alpha\beta\Gamma(-1+2\alpha)\Gamma(-1+2\beta)}{(-1+2\alpha+2\beta)\Gamma(2(-1+\alpha+\beta))}
\end{equation}
and for the $k^{\t{th}}$ one ($k\geq 2$):
\begin{equation}
\int_0^{S_{k}} \t{d}\mu_{k}\, P_{k-1}^{2\alpha} S_{k-1}^{2(\beta-1)+(k-1)(1+2\alpha)}
=P_{k}^{2\alpha} S_{k}^{2(\beta-1)+k(1+2\alpha)}D_{\alpha,\beta,k},
\end{equation}
with
\begin{equation}
D_{\alpha,\beta,k}=\frac{\Gamma(1+2\alpha)\Gamma(1+2(\beta-1)+(k-1)(1+2\alpha))}{\Gamma(1+2(\beta-1)+k(1+2\alpha))}.
\end{equation}
Therefore the total energy reads
\begin{equation}
\mathcal E =p \times \frac{\alpha\beta\Gamma(-1+2\alpha)\Gamma(-1+2\beta)}{(-1+2\alpha+2\beta)\Gamma(2(-1+\alpha+\beta))}\prod_{k=2}^p D_{\alpha,\beta,k}
\end{equation}
where:
\begin{equation}
\prod_{k=2}^p D_{\alpha,\beta,k}=
\frac{\Gamma(1+2\alpha)^{p-1}\Gamma(1+2(\beta-1)+(1+2\alpha))}{\Gamma(1+2(\beta-1)+p(1+2\alpha))}.
\end{equation}
The variance of the estimator $\Delta^2\tilde\bvar=1/N^2\cdot\mathcal{E}/\mathcal N$ can be simplified to:
\begin{equation}
\label{alphabeta}
\Delta^2\tilde\bvar=\frac{1}{N^2}\cdot\frac{\mathcal{E}}{\mathcal N}=\frac{1}{N^2}\cdot p \times \frac{(1-\alpha-\beta)(-1+2\beta+p+2\alpha p)(2\beta+p+2\alpha)}{2(-1+2\alpha)(-1+2\beta)}.
\end{equation}
By direct calculation one may find, that $\alpha,\beta\geq \frac{1}{2}$ which minimize the expression above are
\begin{equation}
\alpha_{\rm min}=\frac{1}{2}+\sqrt{\frac{4p^2+6p+2-4\sqrt{p(1+2p)^2}}{4(p-1)^2}}=\frac{3}{2}+\mathcal O\left(\frac{1}{\sqrt{p}}\right),
\end{equation}
\begin{equation}
\beta_{\rm min}=\frac{1+2p+\sqrt{2}\sqrt{p(1+2p)^2}\sqrt{\frac{p(3+2p)+1-2\sqrt{p(1+2p)^2}}{(p-1)^2}}}{4p+2}=\sqrt{p}+\mathcal O(1).
\end{equation}
Looking for the leading term for large $p$ we may therefore set $\alpha=\frac{3}{2}$, $\beta=\sqrt{p}$ and then:
\begin{equation}
\Delta^2\tilde\bvar=\frac{1}{N^2}\cdot\frac{\mathcal{E}}{\mathcal N}=\frac{p (1+2\sqrt{p})^2\sqrt{p}(4p+2\sqrt{p}-1)}{N^2(8\sqrt{p}-4)}\overset{p\gg 1}{\approx} \frac{2p^3}{N^2}.
\end{equation}

\begin{figure}[t!]
\includegraphics[width=0.9\columnwidth]{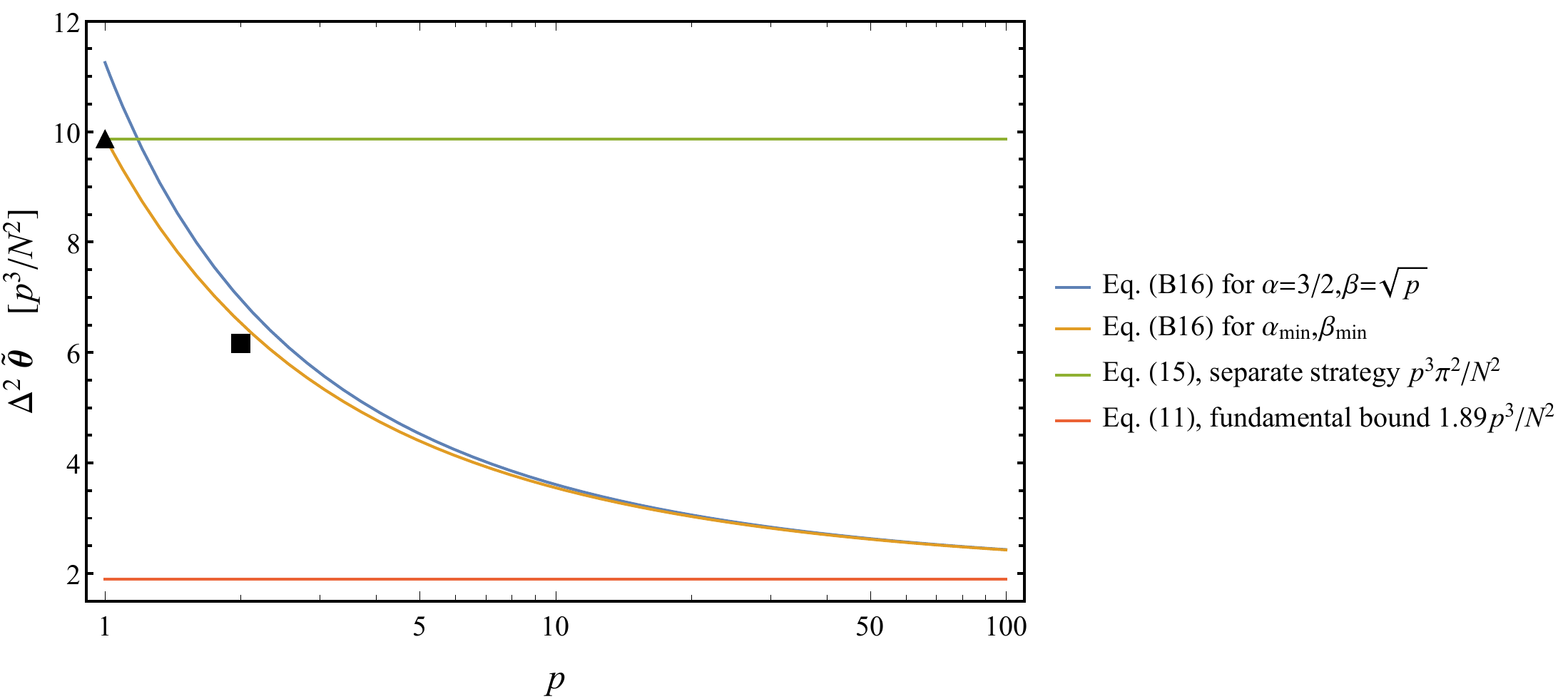}
\caption{Comparison between the cost obtained by the ansatz state, by the optimal separate strategy and the fundamental bound. The lines are plotted as continuous only for better visibility (but they are defined only for natural $p$). For $p=1$ and $p=2$ also analytically found optimal costs are marked respectively by $\blacktriangle$ and $\blacksquare$.}
\label{fig:comp}
\end{figure}

In \figref{fig:comp} we compare how the cost obtained for the ansatz \eqref{alphabeta} changes with increasing $p$ for $\alpha,\beta$ equal respectively $\frac{3}{2},\sqrt{p}$ and $\alpha_{\rm min},\beta_{\rm min}$. As a reference we also plot the cost corresponding to the  separate strategy and the fundamental bound Eq.~(11) (not necessary saturable). We see that the difference in the cost obtained by $\alpha,\beta$ equal $\frac{3}{2},\sqrt{p}$ and $\alpha_{\rm min},\beta_{\rm min}$ is relatively small (almost negligible for $p\geq 10$); therefore, for the simplicity of calculation further we discuss case with $\alpha=\frac{3}{2}$, $\beta=\sqrt{p}$.

It is worth to remind here, that, as mentioned in the main text, the problem given in eqref{eq:jointenergy} has a known exact solution for $p=1,2$ \cite{Gorecki2020pi,li1984particle}, which are respectively:
\begin{align}
\blacktriangle\quad p=1:\quad &f(\mu_1)\propto\sin(\pi \mu_1)&&\to \Delta^2\tilde\bvar=\frac{\pi^2}{N^2}\\
\blacksquare\quad p=2:\quad &f(\mu_1,\mu_2)\propto\sin(\pi \mu_1)\sin(2\pi \mu_2)+\sin(2\pi \mu_1)\sin(\pi \mu_2)&&\to \Delta^2\tilde\bvar=\frac{5\pi^2}{N^2}.
\end{align}
From \figref{fig:comp} one can see that for $p=1,2$ the cost obtained by the ansatz  is very close to the optimal one.

\subsection{The distribution of photons between the arms of the interferometer}
For $\alpha=\frac{3}{2}$, $\beta=\sqrt{p}$, chosen to minimize the total cost of estimation, let us investigate some more features of the state \eqref{appstate}. The distribution of photons in the $i^{\rm th}$ sensing arm is given by the diagonal elements of the corresponding single mode reduced density matrix \eqref{densitym}.
By applying the methods from previous section one may immediately see that it is proportional to:
\begin{equation}
\label{density}
\rho(\mu_i;\mu_i)\propto \mu_i^{3}(1-\mu_i)^{2(2p+\sqrt{p}-2)}.
\end{equation}
The mean number of photons in each sensing arm is therefore given as:
\begin{equation}
\mathbb E[N\mu_i]=\int_0^1 \t{d}\mu_i \rho(\mu_i;\mu_i)N\mu_i=
\frac{4N}{1 + 2 \sqrt{p} + 4 p}\approx \frac{N}{p}.
\end{equation}
From that, the expectation value of the number of photons in the reference arm is:
\begin{equation}
\mathbb E[N-{\textstyle \sum_{i=1}^pN\mu_i}]=N-p\times \mathbb E[N\mu_i] =\frac{(1+2\sqrt{p})N}{1+2\sqrt{p}+4p}\approx \frac{N}{2\sqrt{p}}.
\end{equation}
We see, that it decreases slower with $p$ than the mean number of photons for any single sensing arm; still, in the asymptotic regime $p\to\infty$ it tends to zero and almost all photons are distributed between the sensing arms. Note that analogous situations occurs for the state maximizing QFI \eqref{eq:stateii}. As mentioned in the main text, this is where the advantage of the joint over separate estimation strategy comes from---by using a single reference arm to measure all the phases, more photons remain for the sensing arms.

\begin{figure}[hbt!]
\includegraphics[width=0.6\columnwidth]{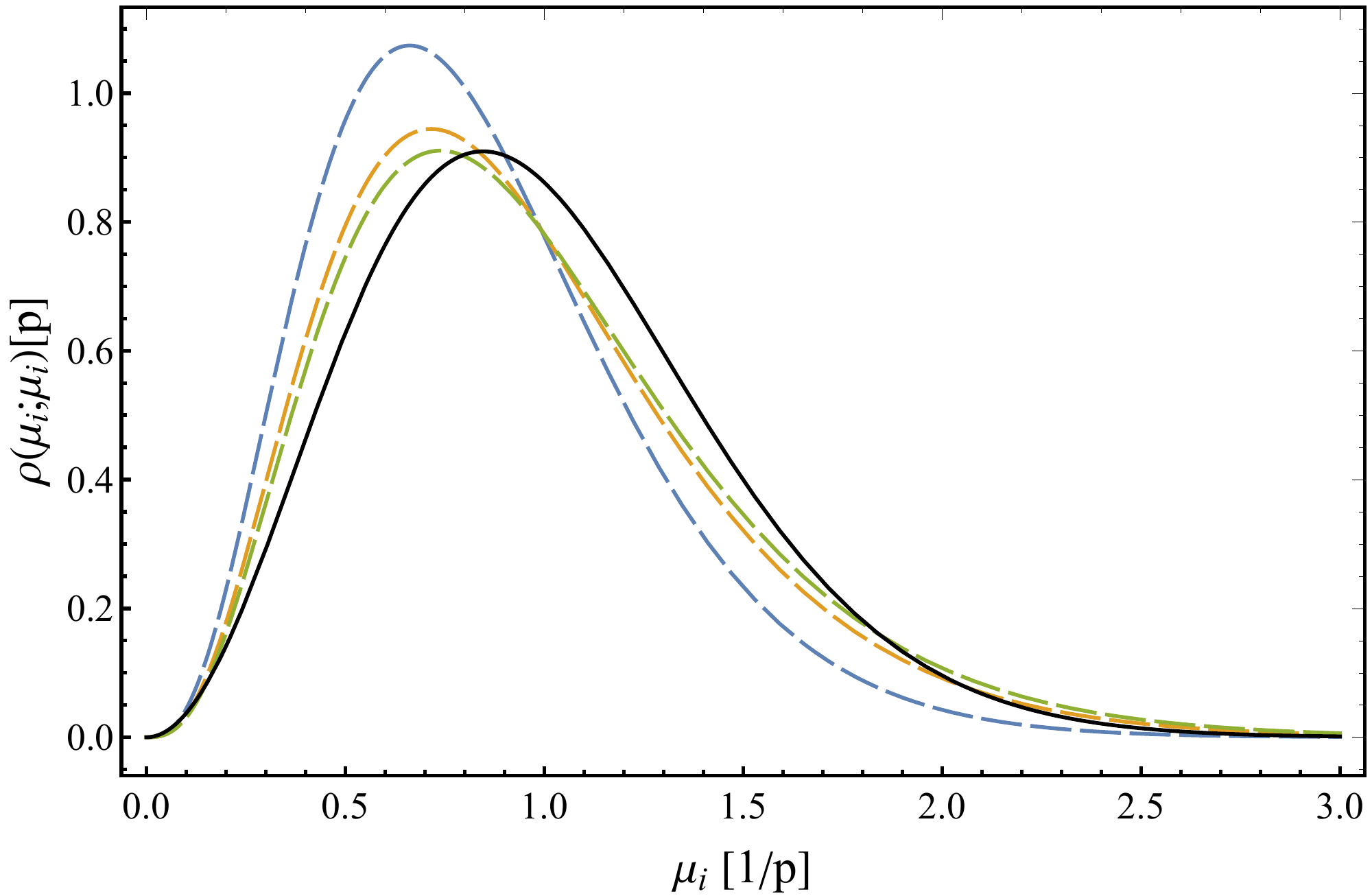}
\caption{See how the distribution of photons in single sensing arm changes with increasing number of phases. The black solid line is normalized square of corresponding Airy function \eqref{Ai2}. The dashed lines correspond to single mode photons number distribution  \eqref{density} for $p$ equal to $10$ (blue), $50$ (orange) and $250$ (green).}
\label{fig:fun}
\end{figure}

An interesting question is if the distributions of photons in different sensing arms are mutually correlated. For the joint distribution  of photons in $i^{\rm th}$ and $j^{\rm th}$ sensing arms we have:
\begin{equation}
\rho(\mu_i,\mu_j;\mu_i,\mu_j)\propto (\mu_i\mu_j)^{3}(1-\mu_i-\mu_j)^{2(2p+\sqrt{p}-4)},
\end{equation}
\begin{equation}
{\rm corr}(N\mu_i,N\mu_j)=\frac{\mathbb E[N\mu_iN\mu_j]-\mathbb E[N\mu_i]\mathbb E[N\mu_j]}{\sqrt{\mathbb E[(N\mu_i)^2]-\mathbb E[N\mu_i]^2}\cdot\sqrt{\mathbb E[(N\mu_j)^2]-\mathbb E[N\mu_j]^2}}=-\frac{4}{4p+2\sqrt{p}-3}\approx -\frac{1}{p},
\end{equation}
which tends to zero with increasing $p$ (unlike for the state maximizing QFI, where numbers of photons in different arms are strongly anti-correlated). Together with the fact that the total cost of estimation for the state \eqref{appstate} is very close to the fundamental bound, it strongly suggests that this state may have a lot in common with the function optimal for single phase estimation with fixed mean number of photons \eqref{eq:g}. To see it directly, we compare the distributions of photons corresponding to both these cases, i.e. \eqref{density} with:
\begin{equation}
\label{Ai2}
|g(\mu)|^2\propto \t{Ai}^2\left(A_0+\frac{2p|A_0|}{3}\mu\right).
\end{equation}
From \figref{fig:fun} one can see, that for big number of phases $p$ both distributions are indeed similar.

\subsection{Superiority of joint measurement for finite number of photons $N$---discrete version of \eqref{eq:state}}
So far we have shown that for the model where discrete variables from the original problem $m_i\in\{0,1,...,N\}$ are replaced with continuous ones $\frac{m_i}{N}\to\mu_i\in[0,1]$, and all the parameters are initially completely unknown $\bvar\in\mathbb R^p$, the state of the form given in \eqref{eq:state} offers a significant advantage:
\begin{equation}
\label{jointcont}
\left[\Delta^2\tilde\bvar\right]^{\rm jointly, continuous}_{\rm ansatz}=\frac{p (1+2\sqrt{p})^2\sqrt{p}(4p+2\sqrt{p}-1)}{N^2(8\sqrt{p}-4)}
\end{equation}
over the optimal separate strategy, for which the cost reads:
\begin{equation}
\label{sepcont}
\left[\Delta^2\tilde\bvar\right]^{\rm separately, continuous}_{\rm optimal}=\frac{p^3\pi^2}{N^2}.
\end{equation}
Following the reasoning from~\cite{Imai2009} (where single phase estimation problem was analyzed) we infer that the results obtained for discrete model converge to the above in the limit $N\to\infty$. Still it is an interesting question to investigate this convergence in more detail for finite number $N$. Below we present the respective numerical analysis.

The output state reads:
\begin{equation}
\ket{\Psi_\bvar^N}=\sum_{\boldsymbol{m}: |\boldsymbol{m}| \leq N} c_{\boldsymbol{m}}e^{i\boldsymbol{m}\cdot \bvar}\ket{\boldsymbol{m}},\quad \bvar\in(-\pi,+\pi]^p.
\end{equation}
Let us consider the covariant cost of the form
\begin{equation}
\label{qucost}
\sum_{i=1}^p(\min\{|\tilde\var_i-\var_i|,2\pi-|\tilde\var_i-\var_i|\})^2,
\end{equation}
which takes into account that two phases $+\pi-\epsilon$, $-\pi+\epsilon$ should be seen as arbitrary close when $\epsilon\to 0$. 
We make this choice in order not to keep the cost as similar as possible to the standard quadratic cost  but at the same time respect the phase periodicity (covariant cost)---other common choice  $\sum_{i=1}^p4\sin^2\left(\frac{\tilde\var_i-\var_i}{2}\right)$~\cite{luis1996,buzek1999,Berry2000},
 is (except of the point $|\tilde\var_i-\var_i|=0$) strictly smaller than the variance and this fact would have an impact on the numerical results and demonstrating the convergence of the results would require going to much higher values of $N$. For such a problem, the minimal cost may be obtained by the covariant measurement:
\begin{equation}
\label{covdisc}
\ket{\chi_{\tilde\bvar}}=\frac{1}{\sqrt{(2\pi)^p}}\sum_{\boldsymbol{m}: |\boldsymbol{m}| \leq N}e^{i\boldsymbol{m}\cdot \tilde\bvar}\ket{\boldsymbol{m}}
\end{equation}
(note that in this case, unlike in continuous model, the measurement is not projective; still it is well defined as $\int_{(-\pi,+\pi]^p}\t{d}\tilde\bvar \ket{\chi_{\tilde\bvar}}\bra{\chi_{\tilde\bvar}}=\openone$).
Then we have:
\begin{equation}
\Delta^2\tilde\bvar=
\max_{\bvar\in\Theta}\int\displaylimits_{\mathbb (-\pi,+\pi]^p} \t{d} \tilde{\bvar}\, |\braket{\chi_{\tilde{\bvar}}}{{\Psi_{\bvar}^N}}|^2 \sum_{i=1}^p(\min\{|\tilde\var_i-\var_i|,2\pi-|\tilde\var_i-\var_i|\})^2
= \sum_{i=1}^p\int\displaylimits_{\mathbb (-\pi,\pi]^p} \t{d} \tilde{\bvar}\, |\braket{\chi_{\tilde{\bvar}}}{{\Psi_{0}^N}}|^2 \tilde\var_i^2.
\end{equation}
Let us now focus on $i^{\text{th}}$ element of above sum:
\begin{multline}
\int\displaylimits_{\mathbb (-\pi,\pi]^p} \t{d} \tilde{\bvar}\, |\braket{\chi_{\tilde{\bvar}}}{{\Psi_{0}^N}}|^2 \tilde\var_i^2=\\
=\frac{1}{(2\pi)^p}\int\displaylimits_{(-\pi,\pi]^p}\t{d} \tilde{\bvar}  \sum_{\boldsymbol{m}}\sum_{\boldsymbol{m}'}c_{\boldsymbol{m}}^*c_{\boldsymbol{m}'}e^{i(\boldsymbol{m}'-\boldsymbol{m})\tilde\bvar}\tilde\var_i^2=
\sum_{\boldsymbol{m}}\sum_{\boldsymbol{m}'}c_{\boldsymbol{m}}^*c_{\boldsymbol{m}'} \left(\prod_{j\neq i} \delta_{m_j,m'_j}\right)\cdot
\begin{cases}
(-1)^{m_i-m_i'}\frac{2}{(m_i-m_i')^2}&{\rm for}\quad m_i\neq m_i',\\
\frac{\pi^2}{3}&{\rm for}\quad m_i=m_i'\\
\end{cases},
\end{multline}
where in the last step discrete Fourier transform of $\tilde\var_i^2$ was applied.

\begin{figure}[hbt!]
\includegraphics[width=1\columnwidth]{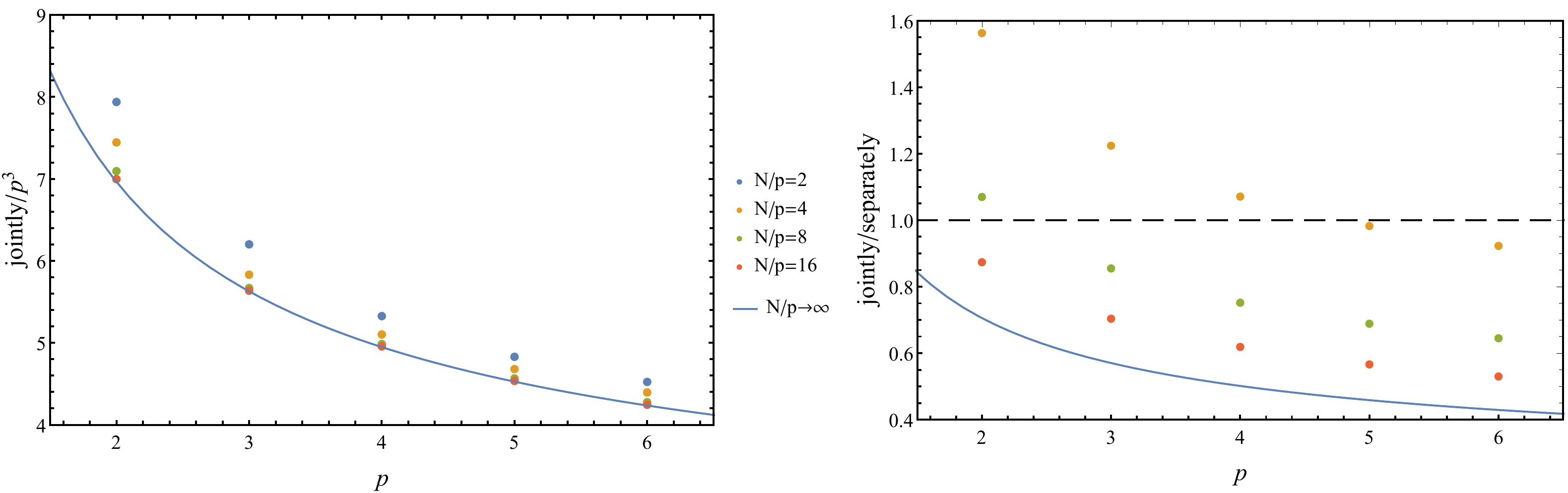}
\caption{In the left panel we present how the cost obtained by applying the Ansazt for discrete model 
$\left[\Delta^2\tilde\bvar\right]^{\rm jointly, discrete}_{\rm ansatz}$ \eqref{joindisc} converges to the one from the continuous model $\left[\Delta^2\tilde\bvar\right]^{\rm jointly, continuous}_{\rm ansatz}$ \eqref{jointcont}
with increasing number of photons $N$. The cost obtained for different number of phases is rescaled in the plot by the factor $1/p^3$ in order to increase the visibility. We see that a good agreement is obtained for $N/p$ of order $10$. 
In the right panel we perform a similar analysis for the advantage of the ansatz over the optimal separate strategy--- 
$\left.\left[\Delta^2\tilde\bvar\right]^{\rm jointly, discrete}_{\rm ansatz}\middle/\left[\Delta^2\tilde\bvar\right]^{\rm separate,discrete}_{\rm optimal}\right.$, 
where the advantage is achieved for a ratio value below $1$ (black dashed line). One can see that for $N/p=16$ the advantage is observed for any number of phases $p$.
In both cases blue lines corresponding to asymptotic limits are plotted as continuous for better visibility (but they are defined only for natural $p$).
}
\label{fig:advantage}
\end{figure}

Now we plug into the above formula the coefficients $c_{\boldsymbol{m}}$ coming from the discretization of \eqref{eq:state}, and arrive at:
\begin{equation}
\label{joindisc}
\begin{split}
\left[\Delta^2\tilde\bvar\right]^{\rm jointly, discrete}_{\rm ansatz}=&p\cdot \sum_{\boldsymbol{m}}\sum_{\boldsymbol{m'}}c_{\boldsymbol{m}}^*c_{\boldsymbol{m}'} \left(\prod_{j\neq i} \delta_{m_j,m'_j}\right)\cdot
\begin{cases}
(-1)^{m_i-m_i'}\frac{2}{(m_i-m_i')^2}&{\rm for}\quad m_i\neq m_i',\\
\frac{\pi^2}{3}&{\rm for}\quad m_i=m_i'\\
\end{cases}\\
&{\rm with}\quad c_{\boldsymbol{m}}\propto f(\boldsymbol{m}/N)=\left(\prod_{i=1}^p m_i/N\right)^{\frac{3}{2}}\left(1-\sum_{i=1}^p m_i/N\right)^{\sqrt{p}},\quad \sum_{\boldsymbol{m}}|c_{\boldsymbol{m}}|^2=1.
\end{split}
\end{equation}
In the left panel of \figref{fig:advantage}, we investigate how \eqref{joindisc} converges to \eqref{jointcont} with the increasing number of photons $N$ for different number of phases $p$. From the plot in may be seen that a good convergence (up to 1\%) is obtained for $N/p$ of the order $10$.

To make the analysis complete, we should also investigate how the \textbf{advantage} of  our strategy over optimal separate one converge to the advantage obtained in continuous case. A certain subtlety should be noted here. The minimal cost obtainable for separate strategy is given as:
\begin{equation}
\left[\Delta^2\tilde\bvar\right]^{\rm separately, discrete}_{\rm optimal}=p\cdot \min_{c_m}\left(
\sum_{m=0}^{N/p}\sum_{m'=0}^{N/p}c_{m}^*c_{m'}\cdot
\begin{cases}
(-1)^{m-m'}\frac{2}{(m-m')^2}&{\rm for}\quad m\neq m',\\
\frac{\pi^2}{3}&{\rm for}\quad m=m'\\
\end{cases}
\right)
\end{equation}
and, in principle, it may be \textbf{smaller} than \eqref{sepcont}. The reason for this is that for finite $N$ the cost depend on the size of $\Theta$ -- which is finite in discrete case, and infinite in continuous. As the consequence, superiority of our ansatz over optimal strategy cannot be judged directly from convergence \eqref{joindisc} to \eqref{jointcont}. Therefore in the right panel of \figref{fig:advantage} we present the convergence of the advantage itself, e.i. how $\frac{\left[\Delta^2\tilde\bvar\right]^{\rm jointly, discrete}_{\rm ansatz}}{\left[\Delta^2\tilde\bvar\right]^{\rm separate,discrete}_{\rm optimal}}$ converge to $\frac{\left[\Delta^2\tilde\bvar\right]^{\rm jointly, continuous}_{\rm ansatz}}{\left[\Delta^2\tilde\bvar\right]^{\rm separate,continuous}_{\rm optimal}}$. Compering both panels one see that convergence of the advantage is indeed slower than convergence of the cost; still, for $N/p=16$ the advantage is observed for any number of phases.

\subsection{Applying to local estimation problem}

Now let us discuss how one may apply above procedure to initial problem, where $\bvar\in\Theta$ (where $\Theta$ is some small finite-sized neighborhood of $\bvar_0$) and standard quadratic cost is considered.
Without loss of generality let as fixed $\bvar_0=0$ (otherwise one could apply proper phase shift $-\var_{0i}$ in each inferetometer's arm) and then choose the smallest cube containing $\Theta$, e.i. $\Theta\in [-d/2,d/2]^p$ (in the opposite to $[-\delta/2,+\delta/2]^p$ laying inside of $\Theta$). Then we apply the covariant measurement \eqref{covdisc}, but each time when the results of measurement point outside of region $[-d/2,d/2]^p$, we choose as the indication of the estimator proper point of the border of this cube. More formally, let as label the measurement outcomes by ${\boldsymbol x}\in (-\pi,+\pi]^p$:
\begin{equation}
\ket{\boldsymbol x}=\frac{1}{\sqrt{(2\pi)^p}}\sum_{\boldsymbol{m}: |\boldsymbol{m}| \leq N}e^{i\boldsymbol{m}\cdot \boldsymbol x}\ket{\boldsymbol{m}}.
\end{equation}
The estimator $\tilde\bvar(\boldsymbol x)=[\tilde\var_1(\boldsymbol x),\tilde\var_2(\boldsymbol x),...,\tilde\var_p(\boldsymbol x)]$ is then:
\begin{equation}
\tilde\var_i(\boldsymbol{x})=
\begin{cases}
x_i\quad&{\rm for}\quad x_i\in [-d/2,+d/2],\\
-d/2\quad&{\rm for}\quad x_i\in (-\pi,-d/2),\\
+d/2\quad&{\rm for}\quad x_i\in (+d/2,+\pi],\\
\end{cases}
\end{equation}
By its construction such procedure for any $\bvar\in [-d/2,+d/2]^p$ (assuming $d\leq \pi$) gives smaller mean cost than the original covariant one discussed previously.

We expect that the difference disappear with increasing $N$, as we expect that propability of poinitng measurement outisde of $[-d/2,+d/2]$ decrease faster than $1/N^2$ (however, we did not prove it analitically). Still, for any finite $N$ results obtained for covariant case bound from above the cost of this strategy.

Therefore, based on the numerical results presented in \figref{fig:advantage} and the analytical bound \eqref{eq:minimaxboundfinal} we may therefore say, that for any non-degenerate $\Theta$, in the limit of large $N$:
\begin{equation}
\forall_{\Theta} 1.89p^3\leq \lim_{N\to\infty}N^2\Delta^2\tilde\bvar\lesssim \frac{p (1+2\sqrt{p})^2\sqrt{p}(4p+2\sqrt{p}-1)}{8\sqrt{p}-4}.
\end{equation}

\renewcommand{\thesection}{\Alph{section}}
\setcounter{theorem}{0}
\renewcommand{\thetheorem}{A\arabic{theorem}}

\end{document}